\documentclass[%
superscriptaddress,
 amsmath,amssymb,
prb,twocolumn%,
]{revtex4-2}

\makeatletter
\def\@hangfrom@section#1#2#3{\@hangfrom{#1#2}#3}%\MakeTextUppercase{#3}}%
\def\@hangfroms@section#1#2{#1#2}%\MakeTextUppercase{#2}}%
\makeatother

\usepackage{siunitx}
\usepackage{graphicx}% Include figure files
\usepackage{dcolumn}% Align table columns on decimal point
\usepackage{bm}% bold math
\usepackage{hyperref}% add hypertext capabilities
\usepackage[%showframe,%Uncomment any one of the following lines to test 
letterpaper,margin=1in]{geometry}

\usepackage{xcolor}
\usepackage{multirow}
\usepackage{placeins}
\usepackage[inline]{enumitem}
\setlist[itemize]{leftmargin=*}
\setlist[enumerate]{leftmargin=*}
\setlist{noitemsep}
\DeclareSIUnit\angstrom{\text {Å}}
\DeclareSIUnit\torr{Torr}

\def\halfohalf{\left(\frac{1}{2},0,\frac{1}{2}\right)}

\begin{document}

% \preprint{APS/123-QED}
%TC:ignore

\title{Complex Magnetic Ordering in Candidate Topological Superconductors}

\author{Purnima P. Balakrishnan}%
 \email{purnima.balakrishnan@nist.gov}
\affiliation{%
NIST Center for Neutron Research, National Institute of Standards and Technology, Gaithersburg, MD 20899, USA.
}%

\author{Hemian Yi}
\affiliation{%
Department of Physics, The Pennsylvania State University, University Park, PA 16802, USA.}%

\author{Zi-Jie Yan}
\affiliation{%
Department of Physics, The Pennsylvania State University, University Park, PA 16802, USA.}%

\author{Wei Yuan}
\affiliation{%
Department of Physics, The Pennsylvania State University, University Park, PA 16802, USA.}%

\author{Andreas Suter}
\affiliation{PSI Center for Neutron and Muon Sciences, 5232 Villigen PSI, Switzerland.}

\author{Christopher J. Jensen}%
\affiliation{%
NIST Center for Neutron Research, National Institute of Standards and Technology, Gaithersburg, MD 20899, USA.
}%

\author{Pascal Manuel}
\affiliation{ISIS Neutron Facility, STFC Rutherford Appleton Laboratory, Chilton, OX11 0QX, Oxfordshire, UK.}

\author{Fabio Orlandi}
\affiliation{ISIS Neutron Facility, STFC Rutherford Appleton Laboratory, Chilton, OX11 0QX, Oxfordshire, UK.}

\author{Takayasu Hanashima}
\affiliation{Neutron Science and Technology Center, Comprehensive Research Organization for Science and Society (CROSS), Tokai, Ibaraki 319-1106, Japan.}

\author{Christy J. Kinane}
\affiliation{ISIS Neutron Facility, STFC Rutherford Appleton Laboratory, Chilton, OX11 0QX, Oxfordshire, UK.}

\author{Andrew J. Caruana}
\affiliation{ISIS Neutron Facility, STFC Rutherford Appleton Laboratory, Chilton, OX11 0QX, Oxfordshire, UK.}

\author{Brian B. Maranville}%
\affiliation{%
NIST Center for Neutron Research, National Institute of Standards and Technology, Gaithersburg, MD 20899, USA.
}%

\author{Zaher Salman}
\affiliation{PSI Center for Neutron and Muon Sciences, 5232 Villigen PSI, Switzerland.}

\author{Thomas Prokscha}
\affiliation{PSI Center for Neutron and Muon Sciences, 5232 Villigen PSI, Switzerland.}

\author{Cui-Zu Chang}
 \email{cxc955@psu.edu}
\affiliation{%
Department of Physics, The Pennsylvania State University, University Park, PA 16802, USA.}%

\author{Alexander J. Grutter}
 \email{alexander.grutter@nist.gov}
\affiliation{%
 NIST Center for Neutron Research, National Institute of Standards and Technology, Gaithersburg, MD 20899, USA.
}%

\date{\today}% It is always \today, today,

\begin{abstract}

The search for chiral topological superconductivity in magnetic topological insulator (TI)-FeTe heterostructures is a key frontier in condensed matter physics, with potential applications in topological quantum computing. The combination of ferromagnetism, superconductivity, and topologically nontrivial surface states brings together the key elements required for chiral Majorana physics. In this work, we examine the interplay between magnetism and superconductivity at the interfaces between FeTe and a series of TI overlayers. In superconducting MnBi$_2$Te$_4$/FeTe, any interfacial suppression of antiferromagnetism must affect at most a few nanometers. On the other hand, (Bi,Sb)$_2$Te$_3$/FeTe layers exhibit near-total suppression of antiferromagnetic ordering. Ferromagnetic Cr$_x$(Bi,Sb)$_{2-x}$Te$_3$ (CBST)/FeTe bilayers exhibit net magnetization in both CBST and FeTe layers, with evidence of interactions between superconductivity and ferromagnetism. These observations identify magnetic TI/FeTe interfaces as an exceptionally robust platform to realize chiral topological superconductivity.

%We find evidence of interfacial magnetic reconstruction and show that while the superconductivity extends into the magnetic layers and influences the magnetism, the presence or absence of superconductivity is completely unrelated to the underlying magnetic state of the constituent materials. Indeed, superconductivity may be observed in samples where the bulk of the FeTe layer is homogenously antiferromagnetic or where the antiferromagnetism is almost completely suppressed. In some superconducting films, therefore, antiferromagnetism suppression must be confined tightly to the interface, affecting a region below the sensitivity limit of approximately 2 nm - 3 nm. These observations identify MTI/FeTe interfaces as an incredibly robust superconducting platform to realize chiral topological superconductivity.

\end{abstract}

\maketitle
%TC:endignore
%\tableofcontents

% \section{Introduction}

The discovery of interface-induced superconductivity in heterostructures of FeTe and the topological insulator (TI) Bi$_2$Te$_3$ revealed a promising candidate for topological superconductivity \cite{he2014two}. Cr-doping the TI to yield ferromagnetic Cr$_x$(Bi,Sb)$_{2-x}$Te$_3$ (CBST)/FeTe bilayers brings together the prerequisites for chiral topological superconductivity: time-reversal symmetry breaking, topologically nontrivial surface states, and superconductivity \cite{yi2024interface, qi2010chiral, doi:10.1126/science.1222360}. Superconductivity also emerges at the interface between FeTe and the antiferromagnetic (AFM) TI MnBi$_2$Te$_4$ (MBT), indicating compatibility with a range of magnetic states \cite{yuan2024coexistence}.% A common thread in these systems may then be the electronic and magnetic states in the FeTe.

FeTe belongs to the same iron chalcogenide family as superconducting FeSe, but is a non-superconducting AFM \cite{hsu2008superconductivity, martinelli2010antiferromagnetism}. Se-doping FeTe yields superconductivity alongside suppression of the AFM order \cite{chen2009electronic, hsu2008superconductivity, tan2013interface, ge2015superconductivity, sales2009bulk, taen2009superconductivity, martinelli2010antiferromagnetism}. It has therefore been suggested that AFM suppression in FeTe plays a key role in the superconducting state, either through competing magnetic interactions or band-bending and charge transfer \cite{yi2023dirac, qin2020superconductivity, owada2019electronic}, but this question remains debated. There is significant overlap between the superconducting and AFM phases in bulk FeTe$_{1-x}$Se$_x$ \cite{martinelli2010antiferromagnetism, liu2010pi, kawasaki2012phase, PhysRevB.80.140511}. However, excess Fe content, which incorporates into interstitial sites, inhibits superconductivity and promotes AFM \cite{PhysRevB.84.224506, PhysRevMaterials.3.114801, viennois2010effect}, so that the low superconducting volume fraction in the overlap region of $x$ cannot exclude phase segregation.

This question is further complicated by the dimensionality change from bulk to thin film heterostructures. FeTe$_{1-x}$Se$_x$ ($x<0.1$) nanoribbons studied with scanning tunneling microscopy (STM) were found to have superconducting edge states at which the AFM order is suppressed, with a non-superconducting AFM interior \cite{ge2020superconductivity}. On the other hand, for monolayer FeTe on a Bi$_2$Te$_3$ crystal, spin-polarized STM indicated the coexistence of a superconducting gap with bi-collinear AFM \cite{manna2017interfacial}. Indeed, there is evidence that in the TI-based systems, the superconductivity results from competition between AFM interactions in the FeTe and ferromagnetic interactions mediated by the topological surface state \cite{yi2023dirac}. In this interpretation, it is these interfacial magnetic interactions, not the FeTe magnetic order, which are central to the emergent superconducting state in the FeTe-based heterostructures.

%At specific \textit{x} values, FeTe$_{1-x}$Se$_x$ has been proposed as a topological superconductor hosting Majorana bound states in magnetic vortex cores\cite{zhang2018observation, wang2018evidence, zhu2020nearly}.

%These FeTe-based superconductors are key candidate platforms for fault-tolerant quantum computing, and while the interplay of magnetism and superconductivity in these platforms is critical to eventual device operation, much remains poorly understood\cite{kitaev2003fault}. 

\begin{table*}[ht!]
    \centering
    \begin{tabular}{|p{0.20\columnwidth}|p{0.35\columnwidth}|p{0.35\columnwidth}|p{0.35\columnwidth}|p{0.4\columnwidth}|}
    \hline 
     Bilayer System & BST/FeTe on~SrTiO$_3$~(001)& CBST/FeTe on~SrTiO$_3$~(001)& MBT/FeTe on~SrTiO$_3$~(001)& Te/FeTe on~SrTiO$_3$~(001)\\
    \hline 
     Overlayer & (Bi,Sb)$_2$Te$_3$ & Cr$_x$(Bi,Sb)$_{2-x}$Te$_3$ & MnBi$_2$Te$_4$ & Te \\
    \hline 
     Overlayer Properties & Non-magnetic topological insulator & Ferromagnetic \mbox{topological} insulator & Antiferromagnetic \mbox{$T_N = 24$ K \cite{otrokov2019prediction, zhang2013topology}} \mbox{topological} insulator & Non-magnetic, \mbox{topologically} trivial \\
    \hline 
 Bilayer Properties & Superconducting \mbox{$T_{c,onset}$ $\approx$ 12.5 K} \mbox{$T_{c,0}$ $\approx$ 10 K} & Superconducting \mbox{$T_{c,onset}$ $\approx$ 12.5 K} \mbox{$T_{c,0}$ $\approx$ 10 K} & Superconducting \mbox{$T_{c,onset}$ $\approx$ 11 K} \mbox{$T_{c,0}$ $\approx$ 9 K} & Non-superconducting \\
     \hline
FeTe AFM & Suppressed & Suppressed & Not suppressed & Not suppressed \\
     \hline
    \end{tabular}
    \caption{Table of properties for each heterostructure studied. Properties listed for individual constituent materials represent the properties of the material in thin film form, but not in the heterostructure.}
    \label{tab:design}
\end{table*}

Fe-content and AFM order in thin films are challenging to probe, so their role in superconducting TI/FeTe heterostructures remains a mystery. Here, we address these questions using thin film neutron diffraction, low-energy muon spin relaxation (LE-$\mu$SR) spectroscopy, and polarized neutron reflectometry (PNR). We study (Bi,Sb)$_2$Te$_3$ (BST)/FeTe, CBST/FeTe,  MBT/FeTe, and Te/FeTe bilayers, where BST is a nonmagnetic TI while CBST and MBT are the prototypical magnetic TIs (MTIs) \cite{chen2009experimental, gong2019experimental, otrokov2019prediction, chang2013experimental, deng2020quantum}. The electrical transport characteristics of these systems have been reported in our prior studies \cite{yi2023dirac, yi2024interface, yuan2024coexistence}, with onset temperatures ($T_{c,onset}$) between 11~K -- 12.5~K and zero-resistivity ($T_{c,0}$) between 9~K -- 10~K. Representative temperature-dependent longitudinal resistance curves are shown in the Supplemental Information. Table \ref{tab:design} summarizes the characteristics of the bilayers studied, while their crystal structures are shown in Figure \ref{fig:AFM}(a-c).

Superconductivity is observed in systems where the AFM is nearly suppressed (BST/FeTe, CBST/FeTe) as well as those with nearly homogeneous AFM order (MBT/FeTe). LE-$\mu$SR indicates that any AFM suppression in both superconducting MBT/FeTe and non-superconducting Te/FeTe bilayers must affect a layer less than 2 nm thick, complicating the relationship between AFM and superconductivity in FeTe. Furthermore, we observe unexpected magnetic order near the interfaces in BST/FeTe and MBT/FeTe as well as a net magnetization within the FeTe. PNR provides direct evidence that the superconducting region is surrounded by ferromagnetic order in CBST/FeTe.  With reported thicknesses of the superconducting region ranging from 6 nm -- 11 nm\cite{yi2024interface, yuan2024coexistence}, our observations reveal interfacial superconductivity overlapping and coexisting with both AFM and ferromagnetism.%, pointing to a singularly promising system in which to pair time-reversal symmetry breaking with topological superconductivity, in keeping with observations of an extremely high upper critical magnetic field \cite{yi2024interface, yuan2024coexistence}.

\section{Magnetic order in FeTe underlayers}

\begin{figure*}
    \centering
    \includegraphics[width=1.0\textwidth]{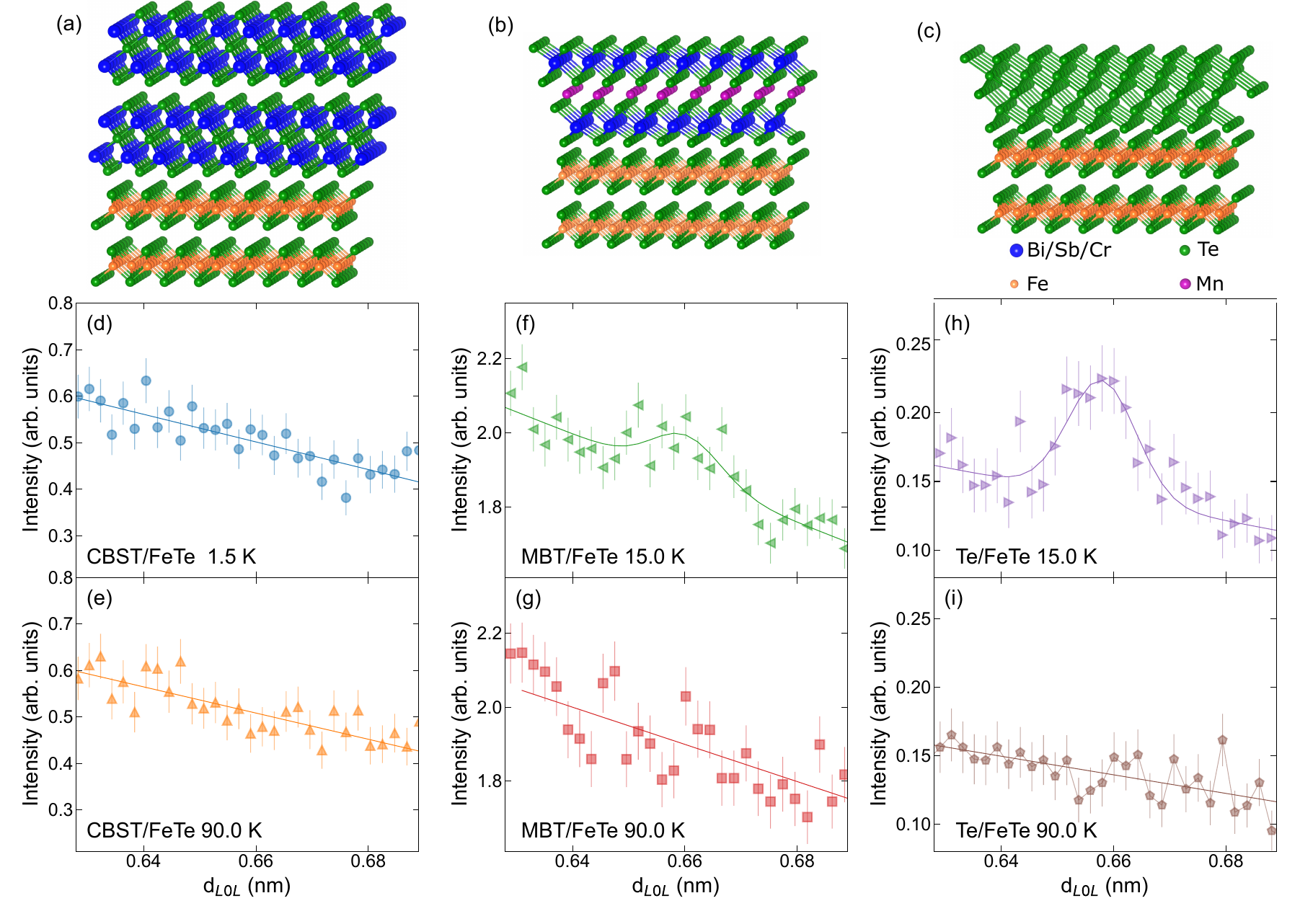}
    \caption{(a-c) Schematics \cite{momma2008vesta} of the (a) ferromagnetic CBST/FeTe, (b) AFM MBT/FeTe, and (c) non-superconducting Te/FeTe bilayer heterostructures. (d-e) A cut through the neutron diffraction data along the (L0L) direction of a representative CBST (15 nm)/FeTe (30 nm) sample at (d) $T$ = 1.5 K and (e) $T$ = 90 K. Cuts are taken at the expected position of the bulk-like ($\frac{1}{2}$ 0 $\frac{1}{2}$) film reflection. Similar cuts are shown for (f-g) an MBT (10 nm) / FeTe (22 nm) sample at (f) $T$ = 15 K and (g) $T$ = 90 K, and for (h-i) a Te (15 nm) / FeTe (30 nm) sample at (h) $T$ = 15 K and (i) $T$ = 90 K. All diffraction curves are plotted alongside fits to the data -- a linear background plus a Gaussian peak, if the fitted area was statistically significant.
    }
    \label{fig:AFM}
\end{figure*}

To determine the magnetic order of the FeTe layers, we performed neutron diffraction measurements on superconducting CBST (15 nm)/FeTe (30 nm), superconducting MBT (10 nm)/FeTe (22 nm), and non-superconducting Te (15 nm)/FeTe (30 nm). Structural peaks for the SrTiO$_3$ (001) substrate and FeTe films are shown in the Supplemental Information. Magnetic peaks should be temperature-dependent, appearing only in the ordered state. Figure \ref{fig:AFM} shows cuts along the [L0L] direction through the FeTe $\halfohalf$ reflection, which is the most intense magnetic diffraction peak associated with bi-collinear AFM order in bulk FeTe. %We note that the TI overlayers are too thin to be measured.

There is no evidence of magnetic diffraction peaks in CBST/FeTe down to 1.5 K (Figs. \ref{fig:AFM}(d-e)). On the other hand, both MBT/FeTe (Figs. \ref{fig:AFM}(f-g)) and Te/FeTe (Figs. \ref{fig:AFM}(h-i)) exhibit statistically significant $\halfohalf$ peaks, appearing at $T$ = 15~K but not at $T$ = 90~K, above the N\'{e}el temperature of bulk FeTe\cite{martinelli2010antiferromagnetism}. The ratio of MBT/FeTe and Te/FeTe peak intensities scale precisely as expected with the FeTe thicknesses of the two samples, as discussed further in the Supplemental Information. We searched for other magnetic diffraction peaks, such as an incommensurate helimagnetic state \cite{PhysRevB.88.165110}, but no evidence of such reflections were found. Therefore, neutron diffraction supports suppressed AFM order in CBST/FeTe and long-range bi-collinear AFM order in both MBT/FeTe and Te/FeTe.

\begin{figure*}
    \centering
    \includegraphics[width=1.0\textwidth]{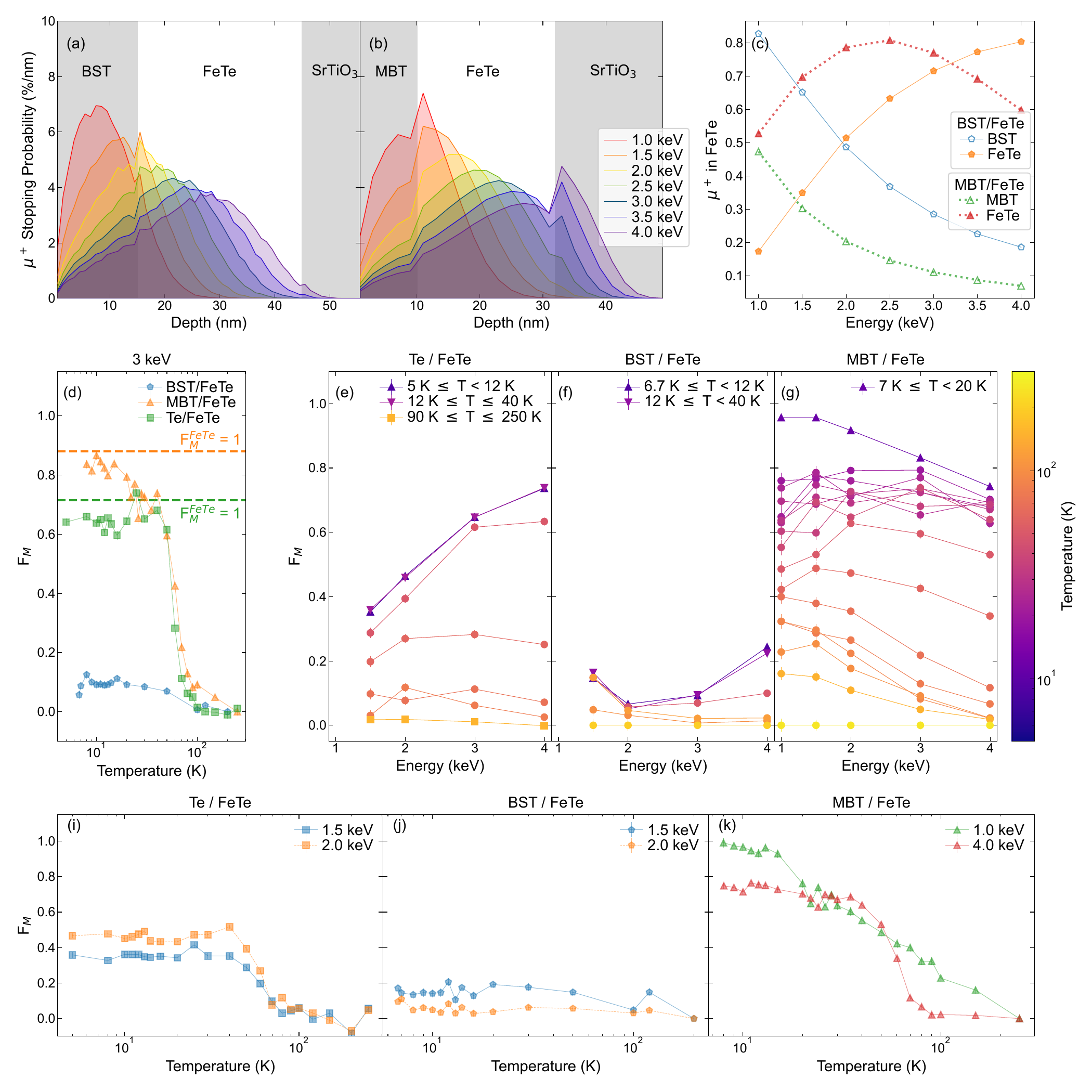}
    \caption{(a) Simulated muon stopping probability as a function of depth expressed as the \% of implanted muons expected to stop per nm for a given bin for the BST (15 nm)/FeTe (30 nm) bilayer. (b) Muon stopping probability for the MBT (10 nm)/FeTe (22 nm) bilayer. (c) The calculated fraction of muons stopping in BST, MBT, and FeTe layers vs. implantation energy. (d) $F_M$ as a function of temperature for BST/FeTe, MBT/FeTe, and Te/FeTe at 3 keV. Dashed lines indicated expected value for $F_M$ if all magnetic layers are 100\% ordered. (e) $F_M$ vs. implantation energy and temperature for Te (15 nm) / FeTe (30 nm), (f) BST (15 nm) / FeTe (30 nm), and (g) MBT (10 nm)/FeTe (22 nm) bilayers. Note that some temperature regimes, indicated in the legend, have been averaged if little or no temperature dependence was noted. The full unbinned dataset is plotted in the Supplemental Information. (i)  $F_M$ vs. temperature for Te (15 nm)/FeTe (30 nm) at 1.5 keV and 2.0 keV. (j)  $F_M$ vs. temperature for BST (15 nm)/FeTe (30 nm) at 1.5 keV and 2.0 keV. (k)  $F_M$ vs. temperature for MBT (10 nm)/FeTe (22 nm) at 1.0 keV and 4.0 keV. 
    }
    \label{fig:MuSR1}
\end{figure*}

Magnetic interactions at the TI/FeTe interface are thought to be central to the superconducting state, but may differ from the rest of the film. We therefore use LE-$\mu$SR to obtain depth-resolved magnetic information. By varying the kinetic energies of implanted muons (1 keV -- 5 keV), we tune the muon implantation depth. LE-$\mu$SR probes the local magnetic field at each implantation site, where each muon ($\mu^+$) spin precesses about the local magnetic field before decaying to yield a positron. The direction of positron emission is asymmetric with respect to the muon spin direction at the moment of decay. Thus, applying a weak transverse magnetic field --  perpendicular to the initial muon polarization -- yields a time-dependent directional asymmetry, which can be described as $A(t) = A_0 e^{-\lambda t}\cos(\omega_L t)$. This allows the extraction of the muon depolarization rate $\lambda$, which, along with the precession frequency $\omega_L$, describes the distribution of local magnetic fields $B$. By tracking the temperature-dependence of the directional asymmetry $A_0(T)$ and normalizing it to a nonmagnetic high-temperature baseline, the fraction of muons which stop in a magnetically ordered film region ($F_M$) may be determined. Combining $F_M$ with TRIM.SP simulations of the muon stopping depth distributions\cite{biersack1984sputtering}, we can probe the layer-dependent magnetically ordered volume fraction. 

Figure \ref{fig:MuSR1} compares LE-$\mu$SR data from BST (15 nm)/FeTe (30 nm), MBT (10 nm)/FeTe (22 nm), and Te (15 nm)/FeTe (30 nm) bilayers. The implantation profile simulations for the BST/FeTe (Fig. \ref{fig:MuSR1}a) and MBT/FeTe (Fig. \ref{fig:MuSR1}b) samples show that each implantation energy probes multiple regions of the heterostructure. While the lowest energies are more sensitive to the TI layers, at 3 keV we primarily probe FeTe for both samples, with minimal contribution from the SrTiO$_3$ substrate (Fig. \ref{fig:MuSR1}(c)). The 3 keV data (Fig. \ref{fig:MuSR1}(d)) reveal increased $F_M$ in Te/FeTe and MBT/FeTe upon cooling from 80 K to 40 K. Based on the halfway point of this transition, we estimate a FeTe $T_N$ of 55 K -- 60 K for both bilayers, corresponding to an approximate composition of Fe$_{1.10}$Te \cite{PhysRevB.88.165110}. This stoichiometry matches the bi-collinear AFM order observed \textit{via} neutron diffraction, though it is near the Fe$_{1.11}$Te Lifshitz point above which Fe$_{1+x}$Te is helimagnetic\cite{PhysRevB.88.165110}.

The dashed lines in Fig. \ref{fig:MuSR1}(d) represent the upper bound of $F_M$ based on the stopping profiles, calculated by assuming that all potentially magnetic layers (FeTe, MBT) are 100\% magnetically ordered. In MBT/FeTe, any drop from this maximum could be explained by AFM suppression in either the FeTe or in the MBT. The assumption that all AFM suppression occurs in a region of FeTe at the interface imposes an upper limit on the thickness of this layer. We find that 7.2\% $\pm$ 0.7\% (MBT/FeTe) or 7.0\% $\pm$ 0.7\% (Te/FeTe) of 2 keV muons are implanted in a nonmagnetic region, limiting any nonmagnetic interfacial FeTe layer to thicknesses of less than 2 nm thick both samples. Thus, the FeTe in both MBT/FeTe and Te/FeTe is close to fully magnetically ordered, with effectively identical nonmagnetic volume fractions for both superconducting and non-superconducting bilayers. We conclude that suppressing the AFM is not sufficient to establish superconductivity.

In BST/FeTe samples, the $F_M$ temperature-dependence is much weaker. LE-$\mu$SR requires that the AFM volume fraction of FeTe must be suppressed by 80\% -- 90\%, supporting the interpretation of neutron diffraction from CBST/FeTe. Superconductivity is therefore clearly compatible with AFM suppression throughout the majority of the FeTe layer. 

\section{Magnetic order in TI overlayers}

We next examine magnetic order within the TI overlayer. Figures \ref{fig:MuSR1}(e,f,g) show the energy and temperature-dependent $F_M$ for Te/FeTe, BST/FeTe, and MBT/FeTe, respectively. Below 100 K, 1.5 keV muons implanted into BST/FeTe yield a higher $F_M$, 0.16 $\pm$ 0.01, than 2.0 keV muons, 0.058 $\pm$ 0.007. This unexpected behavior is highlighted in Figs. \ref{fig:MuSR1}(i) and (j), where Te/FeTe displays the expected trend in which higher energies are more sensitive to the FeTe and consequently exhibit higher $F_M$. This trend is reversed in BST/FeTe, implying the emergence of dilute magnetic order in the BST.

% Approximately 65\% and 50\% of muons stop in the BST overlayer at 1.5 keV and 2 keV implantation energies, respectively (Fig. \ref{fig:MuSR1}(c)). 

For MBT/FeTe, the 1 keV, 1.5 keV, and 2 keV implantation energies exhibit $F_M$ increases upon cooling from 250 K to 90 K. This feature is well above T$_N$ for either MBT or FeTe, and weakens at higher implantation energies. These trends are further illustrated in Fig.\ref{fig:MuSR1}(k), which compares  $F_M$(T) at 1 keV and 4 keV for MBT/FeTe. 

% The $\mu^+$ depolarization rate ($\lambda$) also supports unexpected magnetic transitions in the BST. As shown in Fig. \ref{fig:MuSR2}(d), $\lambda$ for 1.5 keV $\mu^+$s increases by a factor of 3$\times$ - 4$\times$ from 200 K to 100 K, where it saturates until the superconducting $T_c$ of approximately 12 K, at which point it increases by another factor of 2$\times$. Energies that probe deeper into the FeTe layer and are less sensitive to the BST layers, such as 2 keV or 3 keV, do not reproduce this behavior.  Rather, $\lambda$ increases much more slowly upon cooling for both 2 keV and 3 keV, with much broader transitions and a total variance of 2$\times$ - 3$\times$ over all temperatures. Therefore, both BST/FeTe and MBT/FeTe bilayers exhibit unexpected signs of magnetic order at energies which primarily probe the TI overlayer.

A promising explanation for emergent magnetic order in BST and MBT overlayers is Fe-doping. Excess Fe in FeTe$_{1-x}$Se$_x$ has long been thought to suppress superconductivity and strengthen AFM order\cite{PhysRevB.84.224506, PhysRevMaterials.3.114801, viennois2010effect}. Fe incorporation into TIs is also reported to yield a net magnetization, so that the diffusion of excess Fe from the Fe$_{1.1}$Te into the BST or MBT could simultaneously stabilize dilute magnetism in the BST and MBT while promoting interfacial superconductivity\cite{10.1063/1.4788834, 10.1063/1.3549553, kander2023effect}.

\begin{figure*}
    \centering
    \includegraphics[width=1.0\textwidth]{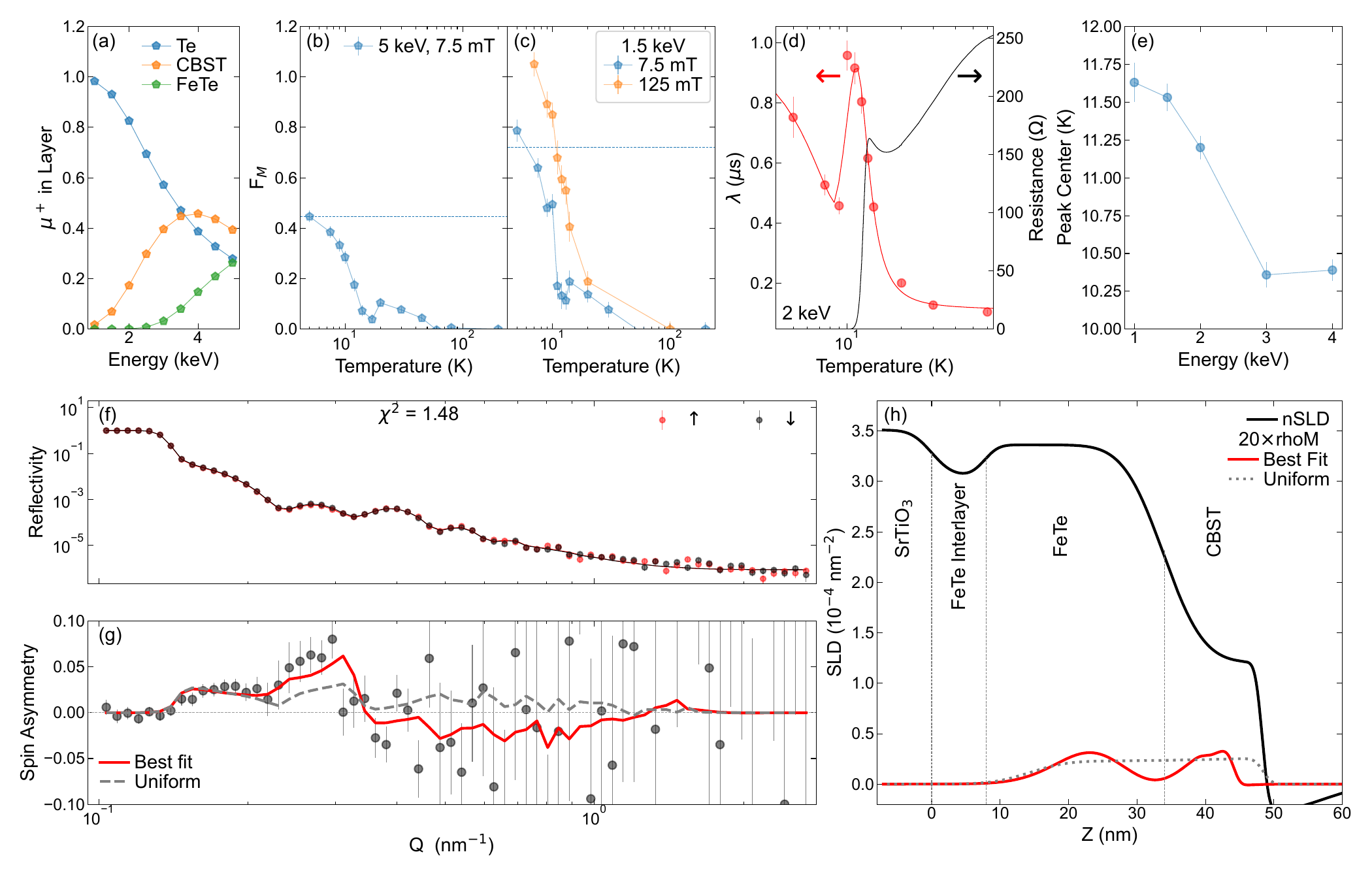}
    \caption{(a) Muon fraction stopping in each layer of Te (25 nm)/CBST (15 nm)/FeTe (10 nm) vs. implantation energy. (b)  $F_M$ vs. temperature for 5 keV muons in Te (25 nm)/CBST (15 nm)/FeTe (10 nm), with a dashed line showing the expected value for 100\% ferromagnetically ordered CBST. Measured in a transverse field of 7.5 mT (c)  $F_M$ vs. temperature for CBST (15 nm)/FeTe (30 nm) at an implantation energy of 1.5 keV, with a dashed line showing the expected value for 100\% ferromagnetically ordered CBST. Measured in transverse fields of 7.5 mT and 125 mT. (d) Muon depolarization rate (red circles) vs. temperature alongside fit to the data using a Lorentzian peak and a magnetic order parameter. Data taken at 2 keV and 7.5 mT. The black line is a typical resistance vs. temperature curve for CBST/FeTe bilayers. (e) Temperature of the $\lambda$ peak vs. muon implantation depth. (f) PNR and (g) spin asymmetry were measured for the CBST (15 nm)/FeTe (30 nm) sample under a 1 T applied magnetic field at 4 K. Solid lines represent theoretical fits to the data. The dashed gray line is the spin-asymmetry of a fit without an interfacial zero-magnetization region. (h) nSLD and mSLD profiles corresponding to the fitted curves in (f) and (g). 
    }
    \label{fig:CBST}
\end{figure*}

% (a) Simulated muon stopping probability as a function of depth expressed as the \% of implanted muons expected to stop per nm for a given bin for a Te (25 nm)/BST (15 nm)/FeTe (10 nm) bilayer. 

All LE-$\mu$SR data presented thus far has focused on non-ferromagnetic TI layers. With measurements on undoped BST/FeTe as a baseline, we evaluate the magnetic ordering in the ferromagnetic Cr-doped TI system. We performed LE-$\mu$SR measurements on trilayer Te (25 nm)/CBST (15 nm)/FeTe (10 nm) and bilayer CBST (15 nm)/FeTe (30 nm). A layer-by-layer breakdown of the expected muon locations for the trilayer samples is shown as a function of implantation energy in Fig. \ref{fig:CBST}(a). The CBST (15 nm)/FeTe (30 nm) implantation profiles are expected to closely track those already shown for (Bi, Sb)$_2$Te$_3$ (15 nm)/FeTe (30 nm).

Figure \ref{fig:CBST}(b) shows $F_M$ for the FeTe (10 nm) trilayer samples under an applied transverse field of 7.5 mT. These samples exhibit a sharp $F_M$ increase below the superconducting $T_c$. The horizontal dashed line represents the expected  $F_M$ from 100\% ferromagnetically ordered CBST and 20\% AFM ordered FeTe, as suggested by the measurements of BST/FeTe. The 5 K data matches the calculated value, indicating fully ordered CBST below $T_c$. Figure \ref{fig:CBST}(c) similarly shows $F_M$ for 1.5 keV muons in the CBST/FeTe bilayers at 7.5 mT, where the calculated  $F_M$ value (horizontal dashed line) agrees with the 5 K data within a 95\% confidence interval.

In Figs. \ref{fig:CBST}(b) and (c), $F_M$ dips by approximately 0.05 near T$_c$ before increasing upon further cooling. Commensurate with this dip in $F_M$, there is a peak in the muon depolarization rate, shown in Fig \ref{fig:CBST}(d). This peak is absent from the BST/FeTe, MBT/FeTe, and Te/FeTe data, as well as examples previously reported LE-$\mu$SR on CBST films\cite{steinke2022magnetic}. Fitting the peak with a Lorentzian function, we plot the peak temperature vs. muon implantation depth and find a shift to lower temperatures for muons probing deeper into the FeTe. To our knowledge, the $F_M$ dip and $\lambda$ peak have no analogue in the literature, and we speculate an origin in the direct spatial overlap between the ferromagnetism and unconventional superconductivity.

In 125 mT applied fields, $F_M$ increases, reaching $F_M$ $\approx$ 1.0 (Fig. \ref{fig:CBST}(c)). That is, all muons stopping in either CBST or FeTe are in close proximity to large static magnetic fields. The LE-$\mu$SR data, therefore, require field-induced magnetic order in the FeTe in the superconducting state. AFM order in CBST/FeTe is substantially suppressed and would not be expected to exhibit a significant field response. Instead, we propose the polarization of excess interstitial Fe, leading to the emergence of a net magnetization in the FeTe at high field. Such behavior explains magnetometry reported in Bi$_2$Te$_3$ (9 nm)/FeTe (140 nm) by Q. L. He et al.\cite{he2015anisotropic}, who found an unexpected enhancement of the net magnetization below the superconducting T$_c$, similar to the lineshapes shown in Fig. \ref{fig:CBST}(b)-(c).

\section{Magnetism at the Interface}

The emergence of a net magnetization from the polarization of interstitial Fe presents an opportunity to probe the deeper origins of superconductivity in TI/FeTe bilayers. Small interfacial changes in Fe-stoichiometry are challenging to detect and quantify through electron microscopy, especially if a thin layer of interfacial FeTe distributes excess Fe throughout the TI cap. Precisely determining the magnetization distribution within the FeTe layer may provide evidence for interfacial Fe depletion in the form of a layer with reduced magnetization.

Therefore, we performed PNR measurements on CBST (15 nm)/FeTe (30 nm). PNR is sensitive to the depth profile of the composition and density of the sample through the nuclear scattering length density (nSLD) and the net in-plane magnetization through the magnetic scattering length density (mSLD), both of which are determined through theoretical fits to the data. Measurements were performed in a 1 T applied magnetic field at 4 K and 13 K. The 4 K spin-dependent neutron reflectivities are shown in Fig. \ref{fig:CBST}(e), while the spin-asymmetry, defined as the difference between the spin-up and spin-down neutron reflectivities normalized by their sum, is shown in Fig. \ref{fig:CBST}(f). Solid lines represent fits to the data. The best-fit nSLD and mSLD depth profiles for the 4 K data are shown in Fig. \ref{fig:CBST}(g). The fitted thicknesses match the target values almost exactly, and there is a region of the FeTe film with reduced nSLD near the FeTe/SrTiO$_3$ interface. This lower nSLD region is consistent with a discontinuous film and has a thickness of ~7.9 nm $\pm$ 0.3 nm, similar to the reported minimum thickness necessary to achieve the zero-resistance state\cite{yi2024interface}.

%\begin{figure*}
%    \centering
    %\includegraphics[width=1.0\textwidth]{Figures/Figure5.pdf}
    %\caption{Magnetic field experienced by implanted $\mu^+$s as a function of temperature in (a) (Bi,Sb)$_2$Te$_3$ (15 nm) / FeTe (30 nm), (b) Te (15 nm) / FeTe (30 nm), (c) MBT (10 nm) / FeTe (22 nm), and (d) CBST (15 nm) / FeTe (30 nm). 
    %}
%    \label{fig:BField}
%\end{figure*}

At 13 K (see supplemental information), the spin splitting is vanishingly small, and consequently the net magnetizations of the CBST and FeTe layers are on the edge of statistical significance. The data are best fit by a net magnetization of 12.4 kA/m $\pm$ 5.8 kA/m in CBST and 4 kA/m $\pm$ 2 kA/m in FeTe. Below $T_c$, the the spin splitting grows, so that the 4 K data are best fit by 6.3 kA/m $\pm$ 2.6 kA/m within the CBST layer and 9.5 kA/m $\pm$ 1.9 kA/m within the FeTe. At both temperatures, the magnetization within a region near the FeTe/CBST interface is effectively zero. Models which prevented an interfacial layer of suppressed magnetization by imposing uniform magnetization on the CBST and FeTe layers (gray dashed lines) were unable to match the spin asymmetry between Q = 0.2 nm$^{-1}$ and Q = 0.3 nm$^{-1}$, as shown by the gray dashed line in Fig. \ref{fig:CBST}(g). Only models with suppressed magnetization at the interface reproduce the spin asymmetry in this region. Further, attempts to fit the data without a magnetized region of the FeTe layer, shown in Fig. \ref{fig:BadPNR}, increase $\chi^2$ to 1.73. 

% , with a fitted value of -6 kA/m $\pm$ 10 kA/m at 13 K and -12 kA/m $\pm$ 6 kA/m

\section{Discussion and conclusions}

The PNR profiles highlight the importance of Fe-stoichiometry in TI/FeTe bilayers. On the FeTe side of the interface, the emergence of a net magnetization at high magnetic field, which increases dramatically below $T_c$, is consistent across the PNR and LE-$\mu$SR. The region of near-zero net magnetization at the CBST/FeTe interface is consistent with a region in which the interstitial Fe has been depleted through diffusion into the TI overlayer. This would also explain the unexpected magnetic orders in BST and MBT overlayers observed \textit{via} LE-$\mu$SR. Here we note that superconductivity coexists either with nearly homogeneous (MBT/FeTe) or substantially suppressed AFM.  Further, the reported thicknesses of the superconducting region in BST/FeTe and CBST/FeTe range from 5 nm -- 10 nm, while our LE-$\mu$SR results show AFM to be suppressed throughout nearly the entire film. These observations raise the question of whether AFM suppression is necessary for superconductivity or merely a byproduct of other interface effects, such as modified Fe-stoichiometry or interactions mediated by the TI surface state\cite{PhysRevMaterials.3.114801, yi2023dirac}.

% He \textit{et al.} also observed an increase in net magnetization below $T_c$ and speculated that the presence of a vortex lattice encouraged these Fe-moments to align parallel to the applied field. However, given that the magnetic signal in that study was dominated by non-superconducting regions far from the interface, this framework is challenging to apply to our samples. Alternative explanations could include RKKY-like exchange coupling between interstitial Fe-sites mediated by the superconductivity.

Lastly, we discuss overlap between the superconductivity and magnetic order. LE-$\mu$SR and PNR support the emergence of a net magnetization within the FeTe. CBST is shown to order ferromagnetically in the superconducting state, while MBT remains AFM. BST and MBT layers show signs of dilute magnetism, which we speculate is induced by Fe interstitials. Thus, in all three superconducting systems, the interfacial superconductivity is surrounded by magnetic ordering. Further, there is evidence of unusual interactions between the superconductivity and magnetism, including enhancement of the net FeTe magnetization below T$_c$ and a combination of transitory $F_M$ suppression with a peak in $\lambda$. These latter two observations have no apparent precedent in the literature.

% in CBST/FeTe, and an unusual peak in  not previously reported in CBST films, and absent from BST/FeTe and MBT/FeTe. Specifically, within the superconducting transition, all $\mu^+$ implantation energies exhibit a sharp feature coinciding with the superconducting transition. Fitting this feature reveals an interplay between $\mu^+$ probing depth and temperature, with the peak appearing at the highest temperature near the interface and shifting to lower temperature at higher $\mu^+$ energies. We speculate that the peak originates in interactions between superconductivity and CBST ferromagnetism\cite{yi2024interface}. 

This work raises questions regarding the underlying mechanism stabilizing superconductivity in TI/FeTe bilayers. AFM order in MBT/FeTe is indistinguishable from that of non-superconducting Te/FeTe, while BST/FeTe and CBST/FeTe exhibit AFM suppression. Thus, AFM suppression may be an incidental outcome of changing Fe-stoichiometry, rather than the sole requirement for superconductivity. Regardless of the underlying origin, the 5 nm -- 10 nm estimated thickness of the superconducting region requires spatial overlap between the emergent superconductivity and net magnetizations in CBST and FeTe, as well as AFM in MBT and FeTe. This is consistent with previously reported transport measurements which observe an upper critical field exceeding the Pauli limit, suggesting unconventional superconductivity which is highly tolerant of magnetic fields\cite{yi2024interface}. Our study demonstrates that the TI/FeTe interface is an incredibly robust system for entwining a topological superconductor with magnetism to yield the time-reversal symmetry breaking required for chiral Majorana physics. 

\section{Experimental Methods}

\subsection{Sample Synthesis}

The magnetic CBST/FeTe and undoped BST/FeTe heterostructures used in this work were grown in two separate commercial MBE chambers (ScientaOmicron) each with a vacuum better than 2$\times$10$^{-10}$ mbar. The insulating 0.5 mm SrTiO$_3$ (100) substrates used for MBE growth are first soaked in hot deionized water (approximately 80 $^\circ$C) for 1.5 hours and then put in an approximately 4.5\% HCl solution for 1 hour. After that, we anneal these substrates at 974 $^\circ$C for 3 hours in a tube furnace with flowing oxygen. Through these treatments, the surface of SrTiO$_3$ (100) substrates is passivated and reconstructed, which makes it suitable for MBE growth of CBST/FeTe and BST/FeTe heterostructures. Next, the heat-treated SrTiO$_3$ (100) substrates are loaded into the MBE chamber and outgassed at 600 $^\circ$C for 1 hour. High-purity Bi (99.9999\%), Sb (99.9999\%), Cr (99.999\%), Fe (99.995\%), and Te (99.9999\%) are evaporated from Knudsen effusion cells. The growth temperatures are 340 $^\circ$C and 210 $^\circ$C for the FeTe and CBST/BST layers, respectively. The growth rate is 0.4 unit cells per minute for the FeTe layer and 0.2 quintuple layers per minute for the CBST and BST layers. Both FeTe and CBST/BST layers are grown in a Te-rich environment to prevent Te deficiency in these heterostructure samples. The MBE growth is monitored using reflection high-energy electron diffraction (RHEED) patterns. 

For MBT samples, the growth temperatures are 340 $^\circ$C and 270 $^\circ$C for the FeTe and MBT layers, respectively. The growth rate is 0.3 unit cells per minute for the FeTe layer and 0.2 septuple layers per minute for the MBT layer. Both are calibrated by measuring the heterostructure thickness using scanning transmission electron microscopy and atomic force microscopy measurements. The Mn effusion cell purity is (99.9998\%), 

\subsection{Neutron Diffraction}

Neutron diffraction measurements were performed using the time of flight Laue diffractometer WISH at the ISIS neutron and Muon Source\cite{chapon2011wish}. The film samples were mounted on a aluminum strip and oriented with the FeTe b axis vertical, giving access to the H0L scattering plane. The FeTe orientation has been chosen such that the {L0L} direction was in optimal flux condition for the $\halfohalf$ reflection. The measurements were performed at 15 K and 90 K with the help of a Oxford instrument $^4$He cryostat. The raw data were analysed with the Mantid software\cite{arnold2014mantid}.

\subsection{Low-Energy Muon Spin Relaxation}

The LE-$\mu$SR measurements are performed using the Low-Energy Muon Facility (LEM) at the Swiss Muon Source, Paul Scherrer Institute, Switzerland\cite{morenzoni2002implantation,Prokscha_2008}. All experiments were performed using two identical 10$\times$10 mm$^{2}$ of the target heterostructure, except for the undoped BST system, in which case only a single piece was used and the incident beam area was reduced accordingly. Samples were mounted with silver paint on a nickel-coated aluminum plate to minimize background effects. The samples were mounted in a helium flow cryostat (CryoVac, Konti), which is capable of maintaining the sample stage temperature within $\pm$0.1 K of the target temperature. Fully polarized muons ($\mu^+$) are accelerated to variable energies to implant them at different depths. Implantation profiles modeled using TRIM.SP  are used to select energies probing different depths\cite{morenzoni2002implantation}. For our heterostructures, the LE-$\mu$SR measurements utilize all or a subset of the implantation energies -- approximately 1.0 keV, 1.5 keV, 2.0 keV, 3.0 keV, 4.0 keV, and 5 keV. The mean implantation depths are presented for the various heterostructures in Figure \ref{fig:MuonDepth} of the supplemental information. The beam transport settings are set to 12.0 kV. Upon implantation into the sample, the muon polarization is oriented within the film plane, and transverse magnetic fields of 7.5 mT and 125 mT are applied perpendicular to the surface of the sample. Measurements are performed upon cooling with multiple energies at each temperature. Before subsequent cooling, the sample is heated to the maximum measurement temperature (at least 200 K) and the magnet is degaussed.
Muons decay to positrons preferentially along their polarization axis, and these positrons are detected using a set of four scintillation detectors. These detectors are positioned to the left, above, to the right, and below the beam axis. For each muon, the time between implantation and decay is recorded, with a time resolution of $\approx$5 ns. The number of positrons detected at each position i is given by

\begin{equation}
    N_i (t)= B_i+N_0 e^{(-t/\tau_{\mu})} (1+ A_i (t))
\end{equation}

where $\tau_\mu$ = 2.2 $\mu$s is the muon lifetime, B$_i$ is the background signal at each detector. In the chosen weak transverse magnetic field experimental setup used for these experiments, the asymmetry A$_i$(t) is the oscillatory signal caused by the muon spin precession. The time spectra for each detector between 0.1 $\mu$s and 10 $\mu$s are simultaneously fitted using the musrfit software\cite{suter2012musrfit} to the following functional form

\begin{equation}
    A_i(t) = A_0e^{-\lambda t} cos(B\gamma_\mu t + \phi_i)
\end{equation}

where B is the average magnetic field experienced by the muon and $\lambda$ is the depolarization rate, corresponding to the width of the field distribution or characterizing spin fluctuations. By subtracting the signal from reflected muons\cite{suter2023low}, the resulting asymmetry can be used to calculate the magnetically ordered volume fraction:

\begin{equation}
    F_M (E,T)=1-\frac{A_0 (E,T)}{A_0 (E,``high T'' )}
\end{equation}

We note here that there are some signs of a hysteretic temperature dependence in the LE-$\mu$SR data. Specifically, while the vast majority of data were taken while cooling the samples, a few data points were taken during warming due to unavoidable time constraints in the allocated beamtime. This yielded the single largest outlier in the measurement, at 120 K on the BST (15 nm)/ FeTe (30 nm) sample. In this case, the 120 K $F_M$ value does not lie between the 100 K and 200 K data, instead matching the 50 K measurement. We cannot at this time speculate regarding the origin of this hysteresis.

\subsection{Polarized Neutron Reflectometry}

% ##############################################,memo="Takayasu HANASHIMA added at 11, Sep, 2024",
% Polarized neutron reflectometry measurements were performed using the SHARAKU instrument at J-PARC. The neutron spins were polarized parallel or antiparallel to the 1 T applied magnetic field, and the spin-dependent reflectivities were measured as a function of the momentum transfer (Q) along the film normal direction. \textbf{More from Takayasu here, wavelength band, reduction, etc.}
Polarized neutron reflectometry measurements were performed using the SHARAKU (BL17)\cite{Takeda_SHARAKU_ChineseJPhys2012} instrument with a horizontal scattering geometry installed
at the Materials and Life Science Experimental Facility (MLF) of the
Japan Proton Accelerator Research Complex (J-PARC). A pulsed neutron
beam was obtained by the spallation reaction at the mercury target where
3 GeV energy proton beam with a repetition rate of 25 Hz was injected.
The neutron beam has white spectrum, which is analyzed by time-of-flight
(TOF) from 0.010 s to 0.040 s, that means from wavelength $\lambda$ = 0.24 nm to
0.88 nm.\cite{Nakajima_MLF_qubs1030009} The neutron spins were polarized parallel or antiparallel to the 1 T applied magnetic field, and the spin-dependent reflectivities were measured as a function of the momentum transfer (Q) along the film normal direction. The data was collected by a two-dimensional position-sensitive detector using a Multi-Wire Proportional Counter (MWPC).\cite{Toh_MWPC_JPhysConf2014} The SHARAKU uses the event data recording method as
a standard control system ``IROHA" and data acquisition (DAQ) one\cite{Nakatani_IROHA2_NOBUGS2016}, and the data conversion and polarization correction from event data to momentum transfer Q was done by software ``Utsusemi"\cite{Inamura_Utsusemi_JPSJ2013}.

Preliminary PNR was performed at the ISIS facility (UK) on the POLREF beamline, and this data was used to inform future sample growth and design the final PNR experiments.

Data were fit using the Refl1D software package, using the DREAM Markov-Chain Monte Carlo algorithm, which is implemented in the BUMPS Python package for model optimization and uncertainty estimation. 

\section*{Author Contributions}

PPB led the project. Samples were fabricated by HY, ZJY, WY, and CZC. LE-$\mu$SR measurements were performed and analyzed by PPB, AJG, AS, ZS, and TP. Neutron diffraction was performed and analyzed by AJG, PPB, BBM, PM, and FO. Neutron reflectometry was performed and analyzed by PPB, AJG, CJJ, TH, CJK, and AJC. The manuscript was written by AJG and PPB with input from all authors. The study was conceived, designed, and coordinated by PPB, AJG, and CZC.

\begin{acknowledgments}
The work done at Penn State is supported by the DOE grant (DE-SC0023113) and the Penn State MRSEC for Nanoscale Science (DMR-2011839). C. -Z. C. acknowledges the support from the Gordon and Betty Moore Foundation’s EPiQS Initiative (Grant GBMF9063 to C. -Z. C.). 
% ##########,memo="Takayasu HANASHIMA added at 11, Sep, 2024",
The neutron experiments on beamline SHARAKU (BL17) at the Materials and Life Science Experimental Facility of the J-PARC were performed under proposal numbers 2023B0108. The authors acknowledge the Science and Technology Facility Council (STFC) for the provision of beam time on the WISH diffractometer at the ISIS facility (UK). The LE-$\mu$SR experiments were performed at the Swiss Muon Source S$\mu$S, Paul Scherrer Institute, Villigen, Switzerland. The raw data of the neutron diffraction data can be found at \href{https://doi.org/10.5286/ISIS.E.RB2320064}{https://doi.org/10.5286/ISIS.E.RB2320064}.
Preliminary PNR was performed at the ISIS facility (UK) on the POLREF beamline, and this data was used to inform future sample growth and design the final PNR experiments. The raw data for that PNR work can be found at \href{https://doi.org/10.5286/ISIS.E.RB2200030-1}{https://doi.org/10.5286/ISIS.E.RB2200030-1}. Research performed in part at the NIST Center for Nanoscale Science and
Technology. Certain commercial equipment, instruments, software, or materials are identified in this paper in order to specify the experimental procedure adequately. Such identifications are not intended to imply recommendation or endorsement by NIST, nor it is intended to imply that the materials or equipment identified are necessarily the best available for the purpose.

\end{acknowledgments}

\clearpage

\onecolumngrid

\section{Supplemental Information}
\renewcommand{\thefigure}{S\arabic{figure}}
\setcounter{figure}{0}

\subsection{X-ray and Neutron Diffraction}

\begin{figure*}
    \centering
    \includegraphics[width=1.0\textwidth]{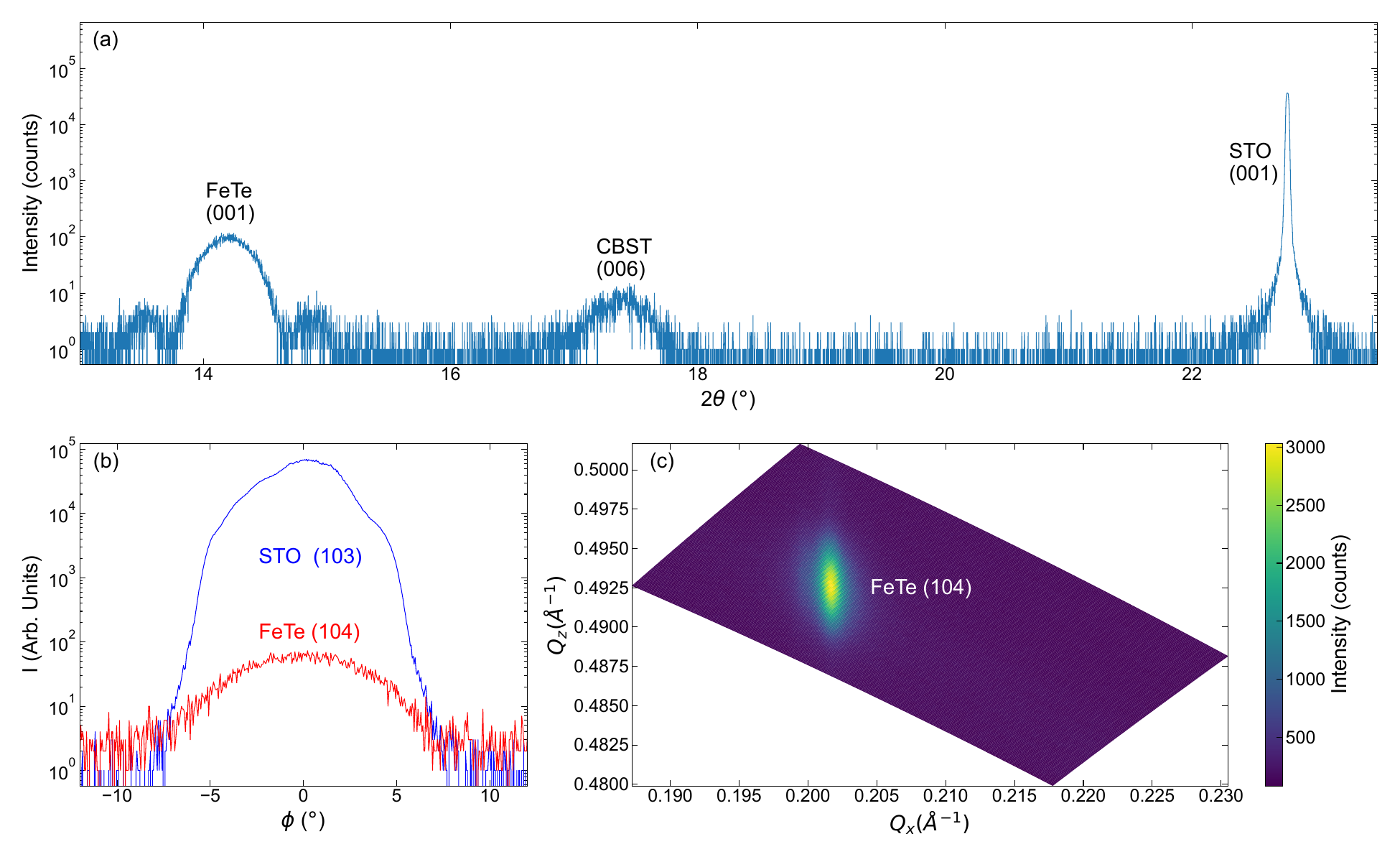}
    \caption{(a) x-ray diffraction symmetry scan of a representative CBST (15 nm)/ FeTe (30 nm) bilayer. (b) $\phi$-scan of the SrTiO$_3$ (103) and FeTe (104) reflections of the same sample. (c) Reciprocal space map of the FeTe (104) reflection.
    }
    \label{fig:xrd}
\end{figure*}

In order to determine the relative orientation of the structural lattices prior to the neutron diffraction experiments, we performed x-ray diffraction measurements on a representative CBST (15 nm)/FeTe (30 nm) bilayer. Figure \ref{fig:xrd}(a) plots a symmetric $\theta-2\theta$ scan along the growth axis, revealing the FeTe (001), CBST (006), and SrTiO$_3$ (001) structural peaks, indicating excellent alignment of the crystal structures along the out-of-plane direction. To confirm the in-plane alignment of the FeTe peaks, we performed $\phi$ scans through the SrTiO$_3$ (103) and FeTe (104) diffraction peaks, Figure \ref{fig:xrd}(b), and a reciprocal space map of the FeTe (104) reflection. As expected, the in-plane [100] direction of the FeTe film is well aligned with the SrTiO$_3$ [100]. The FeTe lattice constants indicate a relaxed film, with \textit{a} $\approx$ 3.81 $\AA$ and \textit{c} $\approx$ 6.26 $\AA$. This is in excellent agreement with measurements on bulk FeTe which report \textit{a} = 3.82 $\AA$ and \textit{c} $\approx$ 6.27 $\AA$\cite{finlayson1956some}. We may, therefore, be confident in the quality and orientation of the FeTe crystal when evaluating the magnetic phases with neutron diffraction.

\begin{figure*}
    \centering
    \includegraphics[width=1.0\textwidth]{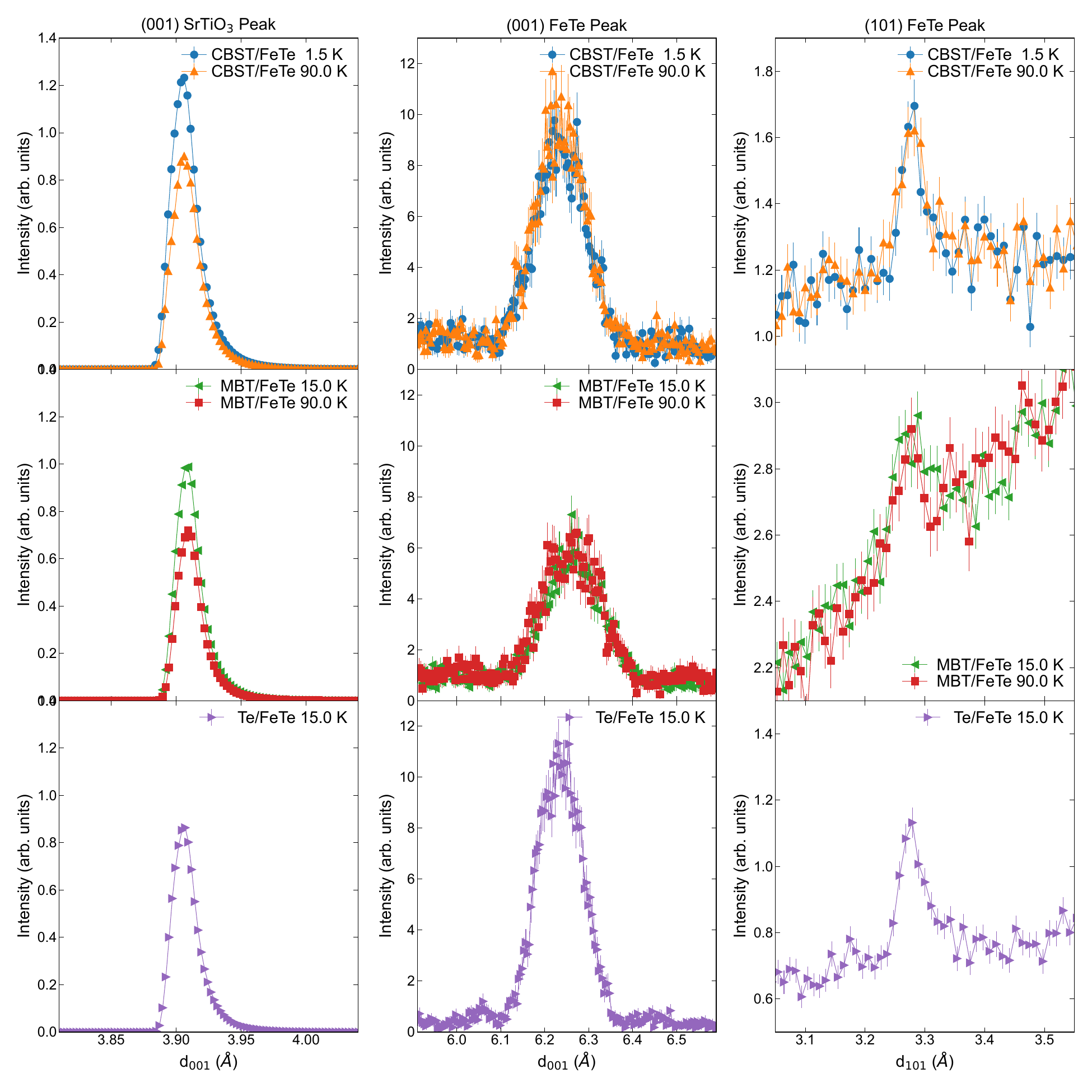}
    \caption{Temperature-dependent neutron diffraction measurements of the SrTiO$_3$ (001), FeTe (001), and FeTe (101) Bragg reflections.
    }
    \label{fig:WISH_nuc}
\end{figure*}

For reference, we reproduce several neutron diffraction peaks which were used to determine the lattice orientation and expected position of the FeTe ($\frac{1}{2}$, 0, $\frac{1}{2}$) reflection. Figure \ref{fig:WISH_nuc} plots the temperature-dependent (001) SrTiO$_3$ structural diffraction peaks for CBST (15 nm)/FeTe (30 nm), MBT (10 nm)/FeTe (22 nm), and Te (15 nm)/FeTe (30 nm) bilayers, all grown on (001) SrTiO$_3$. As expected, the second-order structural phase transition increases the (001) SrTiO$_3$ peak intensity as the sample is cooled. Here we note that each peak type has been scaled relative to other peaks of the same type based on the integrated beam current on the sample, so that, for example, all the (001) SrTiO$_3$ peaks are comparable with each other. However, the (001) SrTiO$_3$ peak intensity is not directly comparable to the (001) and (101) FeTe peak intensities. We further note that all films had approximately the same cross-sectional area, so the intensity differences between peaks of the same type most likely originate from thickness and quality variation among the films. That being said, the low-temperature intensities of the SrTiO$_3$ peaks vary from approximately 0.9 - 1.25, such that a change of 20\% in FeTe peak intensity is likely not significant.

The nuclear FeTe peaks are temperature-independent. Specifically, we show the FeTe (001) and (101)-reflections positioned at the expected locations based on the x-ray diffraction measurements. We conclude, therefore, that the orientation of the crystal lattice is known and well-defined for the purposes of neutron diffraction, and that the SrTiO$_3$ structural transition does not significantly affect the film. As expected, the peak FeTe peak intensities are reduced for the thinner MBT (10 nm)/FeTe (22 nm) sample, but no significant conclusions can be drawn from the intensities beyond this observation.

To analyze the magnetic neutron diffraction peaks shown in the main text, we normalized the data for incident neutron current and fit the ($\frac{1}{2}$, 0, $\frac{1}{2}$) reflection to a Gaussian peak shape using a Markov chain Monte-Carlo (MCMC) algorithm \cite{kienzle2018bumps} for uncertainty estimation yields low-temperature peak areas of 0.000 $\pm$ 0.008, 0.018 $\pm$ 0.006, and 0.029 ± 0.006 for CBST/FeTe, MBT/FeTe, and Te/FeTe, respectively. A second measurement of the Te/FeTe sample using a different sample orientation with higher resolution yields a peak area of 0.0133 $\pm$ 0.002 (Fig \ref{fig:AFM}(h)-(i)), but cannot be directly compared to the other areas due to differences in beam condition, scattering geometry and in the neutron wavelength used. Among comparable scans, the ratio of MBT/FeTe and Te/FeTe peak intensities scales precisely as expected, given the FeTe thicknesses of the two samples. There is no evidence of a ($\frac{1}{2}$, 0, $\frac{1}{2}$) peak in the measurements performed at $T$ = 90 K, above the known bulk FeTe N\'{e}el temperature. 

%Therefore, neutron diffraction supports suppressed antiferromagnetic order in the superconducting CBST/FeTe film and long-range bi-collinear antiferromagnetic order in superconducting MBT/FeTe and metallic Te/FeTe.

% or the helimagnetic \textbf{k} = (0.386, 0, $\frac{1}{2}$) phase observed in iron-rich Fe$_{1+x}$Te above $x \approx 11\%$\cite{PhysRevB.84.064403}.

%Temperature-dependent examples of LE-$\mu$SR asymmetry vs. time histograms are shown for (j) (Bi,Sb)$_2$Te$_3$ (15 nm) / FeTe, (k) MBT (10 nm) / FeTe (22 nm), and (l) Te (15 nm) / FeTe (30 nm) samples, respectively. Error bars represent $\pm$ 1 standard deviation.

\subsection{\texorpdfstring{$\mu^+$ Implantation Depths}{Muon Implantation Depths}}

\begin{figure*}
    \centering
    \includegraphics[width=1.0\textwidth]{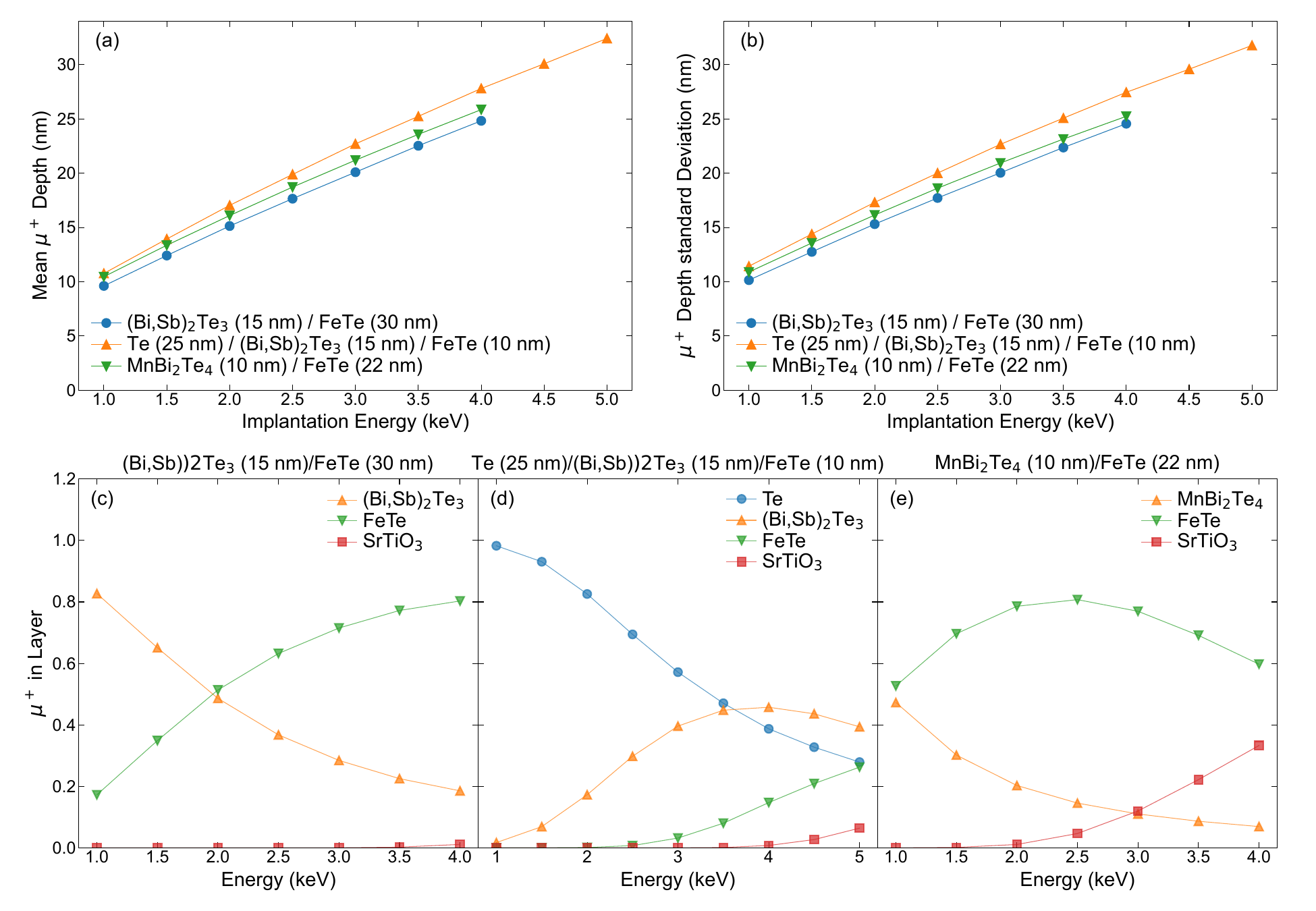}
    \caption{(a) Mean $\mu^+$ implantation depth and (b) standard deviation of the $\mu^+$ implantation depth for BST (15 nm)/FeTe (30 nm), Te (25 nm)/BST (15 nm)/FeTe (10 nm), and MBT (10 nm)/ FeTe (22 nm) samples. (c) Fraction of implanted $\mu^+$ stopping in each layer for the BST (15 nm)/FeTe (30 nm) sample, as well as (d) Te (25 nm)/BST (15 nm)/FeTe (10 nm), and (e) MBT (10 nm)/ FeTe (22 nm).
    }
    \label{fig:MuonDepth}
\end{figure*}

From the $\mu^+$ implantation simulations shown in the main text, it is possible to determine the expected mean and standard deviation
of the $\mu^+$ implantation depth. These calculations are shown in Figure \ref{fig:MuonDepth} for BST (15 nm)/FeTe (30 nm), Te (25 nm)/BST (15 nm)/FeTe (10 nm), and MBT (10 nm)/ FeTe (22 nm) samples. Interestingly, despite the differences in thickness and layering, there is very little change in the average muon implantation depth across samples, especially at implantation energies of 3 keV and below. The same is true of the muon depth standard deviation, which is always extremely close to the mean probing depth.

In Figure \ref{fig:MuonDepth}(c)-(e), we reproduce the calculated fraction of $\mu^+$ stopping in each layer for these same sample geometries.

\subsection{\texorpdfstring{Full LE-$\mu$SR experimental results}{Full LE-muSR experimental results}}

While the main text discussed a subset of the LE-$\mu$SR measurements in detail, we present here the unabridged fitting results from the LE-$\mu$SR analysis. Below, we show the temperature- and implantation energy-dependent values of  $F_M$, $\lambda$, and $B$ for all the presented datasets. For ease of visualization, we present some datasets twice, with both linear and logarithmic temperature scales.

\begin{figure*}
    \centering
    \includegraphics[width=1.0\textwidth]{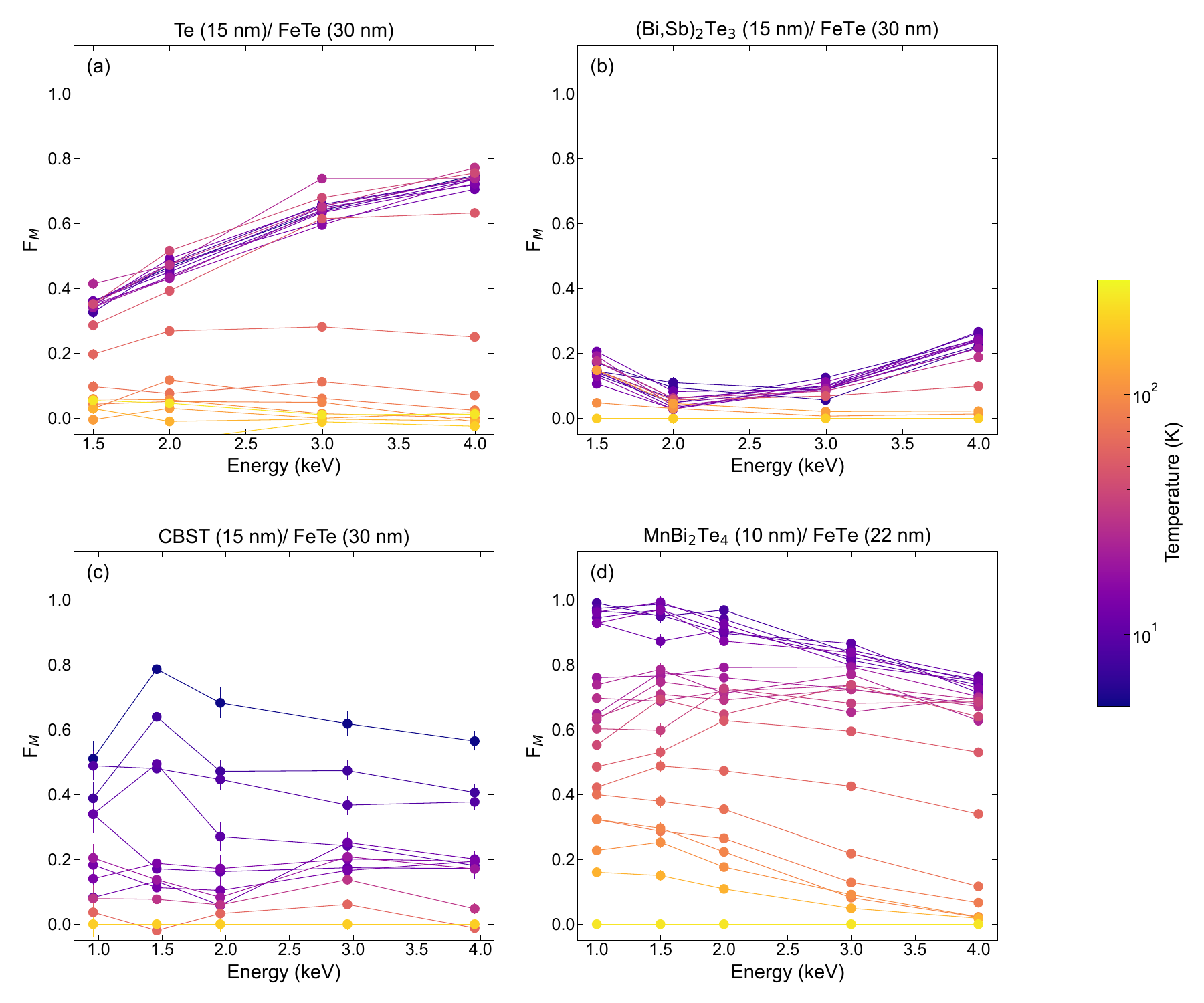}
    \caption{(a) Temperature- and energy-dependent  $F_M$ values extracted from LE-$\mu$SR measurements for Te (15 nm)/FeTe (30 nm), (b) BST (15 nm)/FeTe (30 nm), (c) MBT (10 nm)/FeTe, and (d) CBST (15 nm)/FeTe (30 nm) bilayers. Error bars represent $\pm$1 standard deviation. All measurements shown were formed in an applied external magnetic field of approximately 7.5 mT.
    }
    \label{fig:MuonFMFullvE}
\end{figure*}

\begin{figure*}
    \centering
    \includegraphics[width=1.0\textwidth]{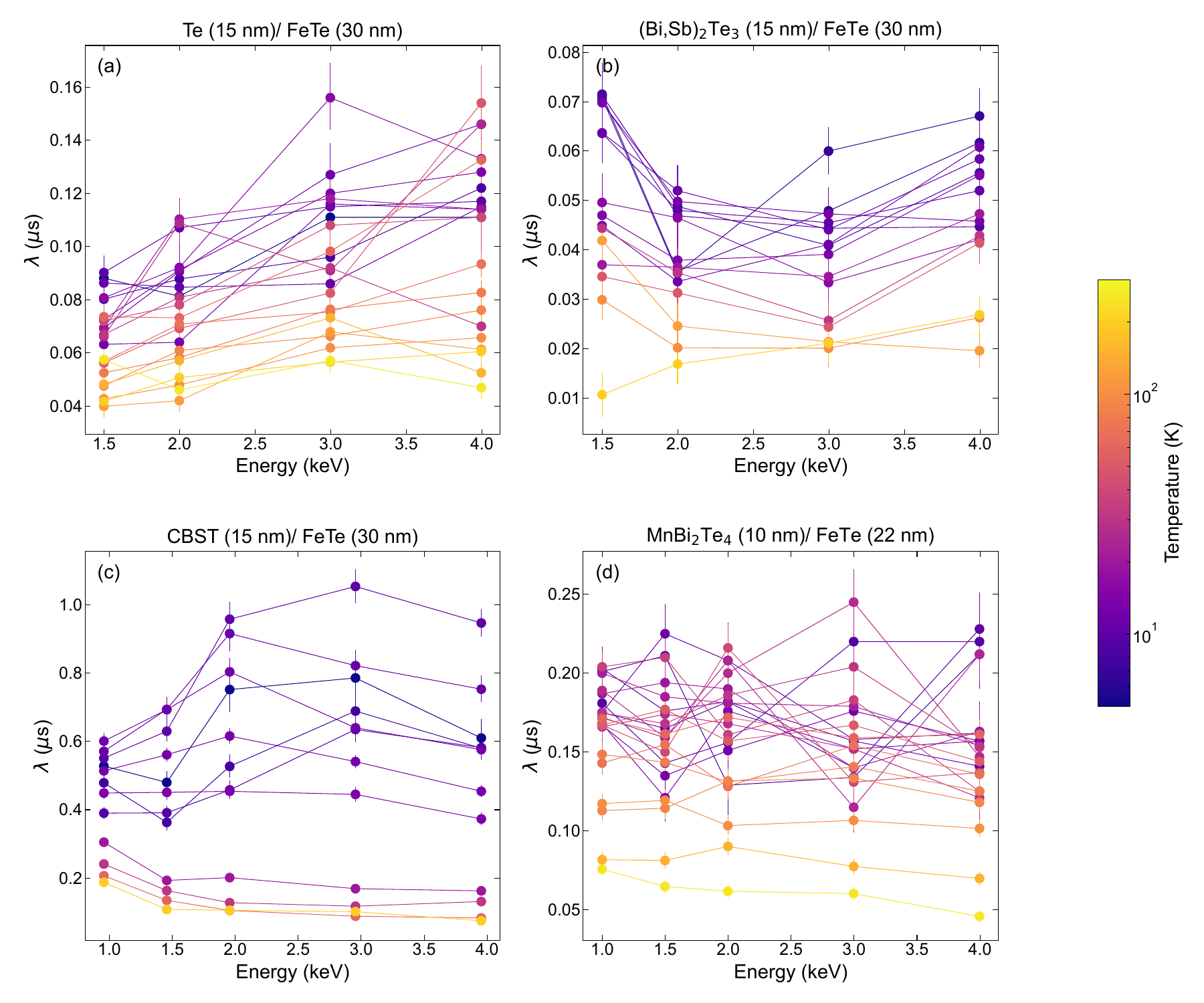}
    \caption{(a) Temperature- and energy-dependent $\lambda$ values extracted from LE-$\mu$SR measurements for Te (15 nm)/FeTe (30 nm), (b) BST (15 nm)/FeTe (30 nm), (c) MBT (10 nm)/FeTe, and (d) CBST (15 nm)/FeTe (30 nm) bilayers. Error bars represent $\pm$1 standard deviation. All measurements shown were formed in an applied external magnetic field of approximately 7.5 mT.
    }
    \label{fig:MuonLbamdaFullvE}
\end{figure*}

\begin{figure*}
    \centering
    \includegraphics[width=1.0\textwidth]{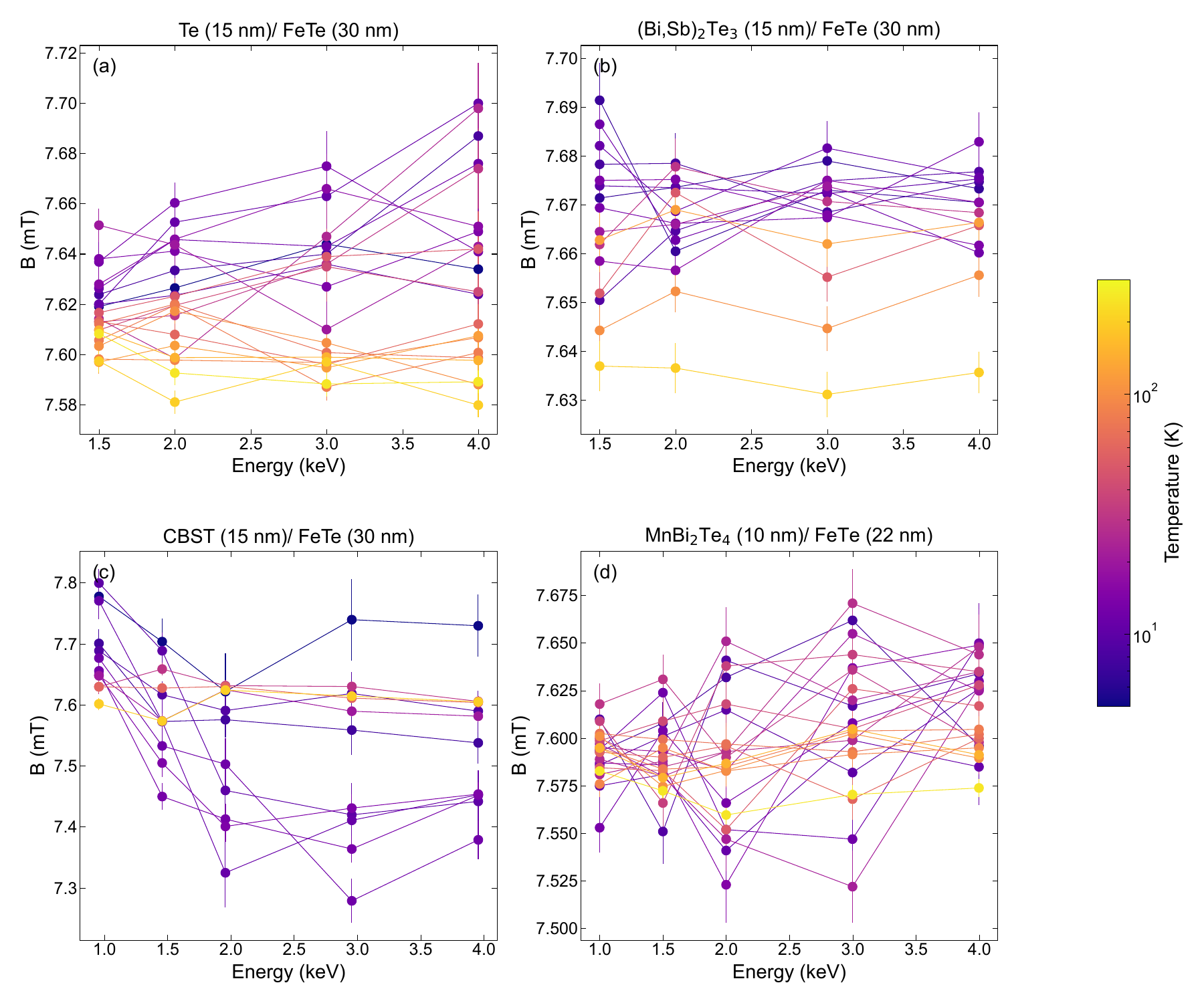}
    \caption{(a) Temperature- and energy-dependent magnetic field (B) values extracted from LE-$\mu$SR measurements for Te (15 nm)/FeTe (30 nm), (b) BST (15 nm)/FeTe (30 nm), (c) MBT (10 nm)/FeTe, and (d) CBST (15 nm)/FeTe (30 nm) bilayers. Error bars represent $\pm$1 standard deviation. All measurements shown were formed in an applied external magnetic field of approximately 7.5 mT.
    }
    \label{fig:MuonBFullvE}
\end{figure*}

\begin{figure*}
    \centering
    \includegraphics[width=1.0\textwidth]{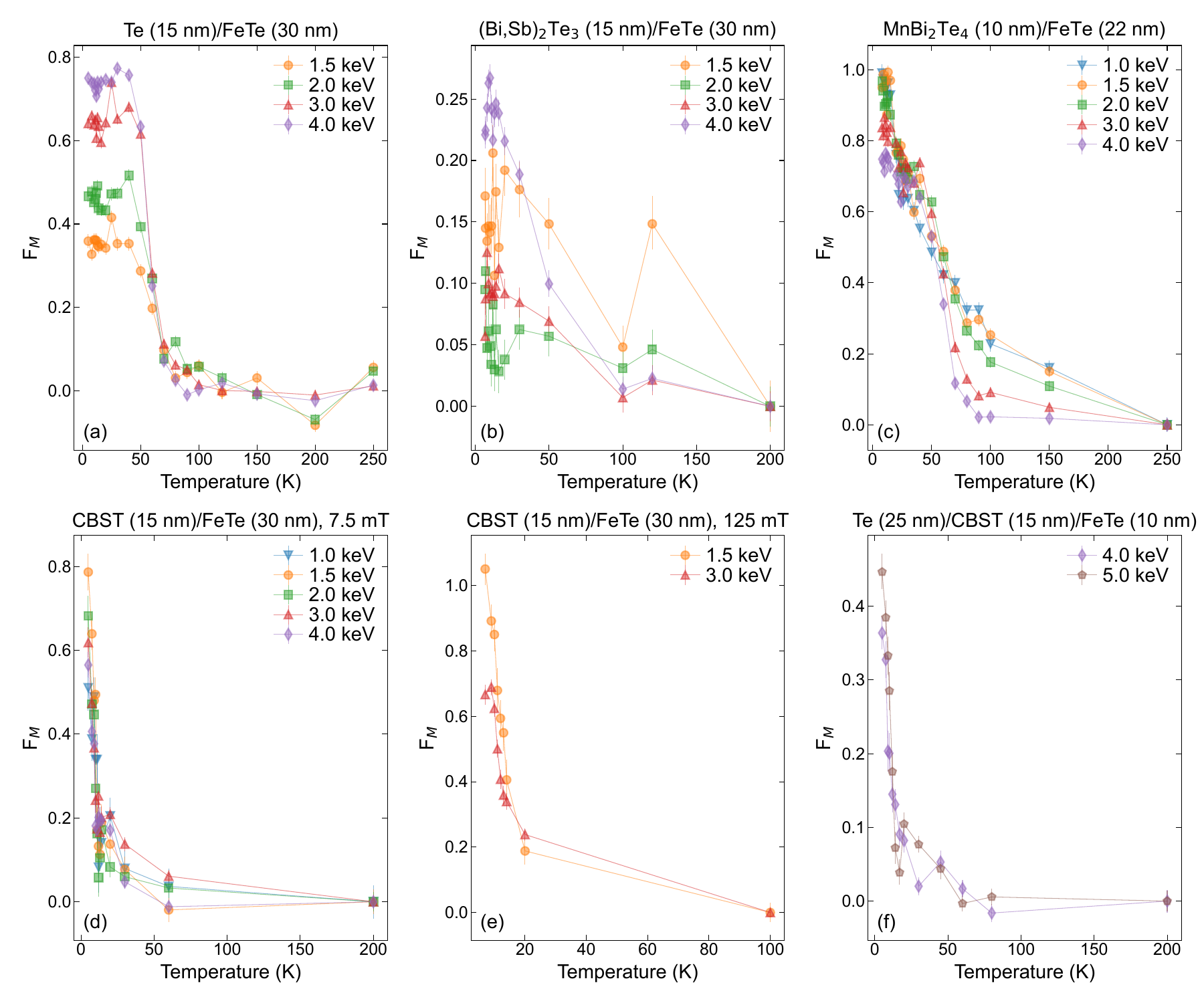}
    \caption{(a) Temperature- and energy-dependent  $F_M$ values extracted from LE-$\mu$SR measurements for Te (15 nm)/FeTe (30 nm) at 7.5 mT, (b) BST (15 nm)/FeTe (30 nm) at 7.5 mT, (c) MBT (10 nm)/FeTe at 7.5 mT, (d) CBST (15 nm)/FeTe (30 nm) at 7.5 mT, (e) CBST (15 nm)/FeTe (30 nm) at 125 mT, and (f) Te (25 nm)/CBST (15 nm)/FeTe (10 nm) at 7.5 mT. Error bars represent $\pm$1 standard deviation.
    }
    \label{fig:MuonFMFull}
\end{figure*}

\begin{figure*}
    \centering
    \includegraphics[width=1.0\textwidth]{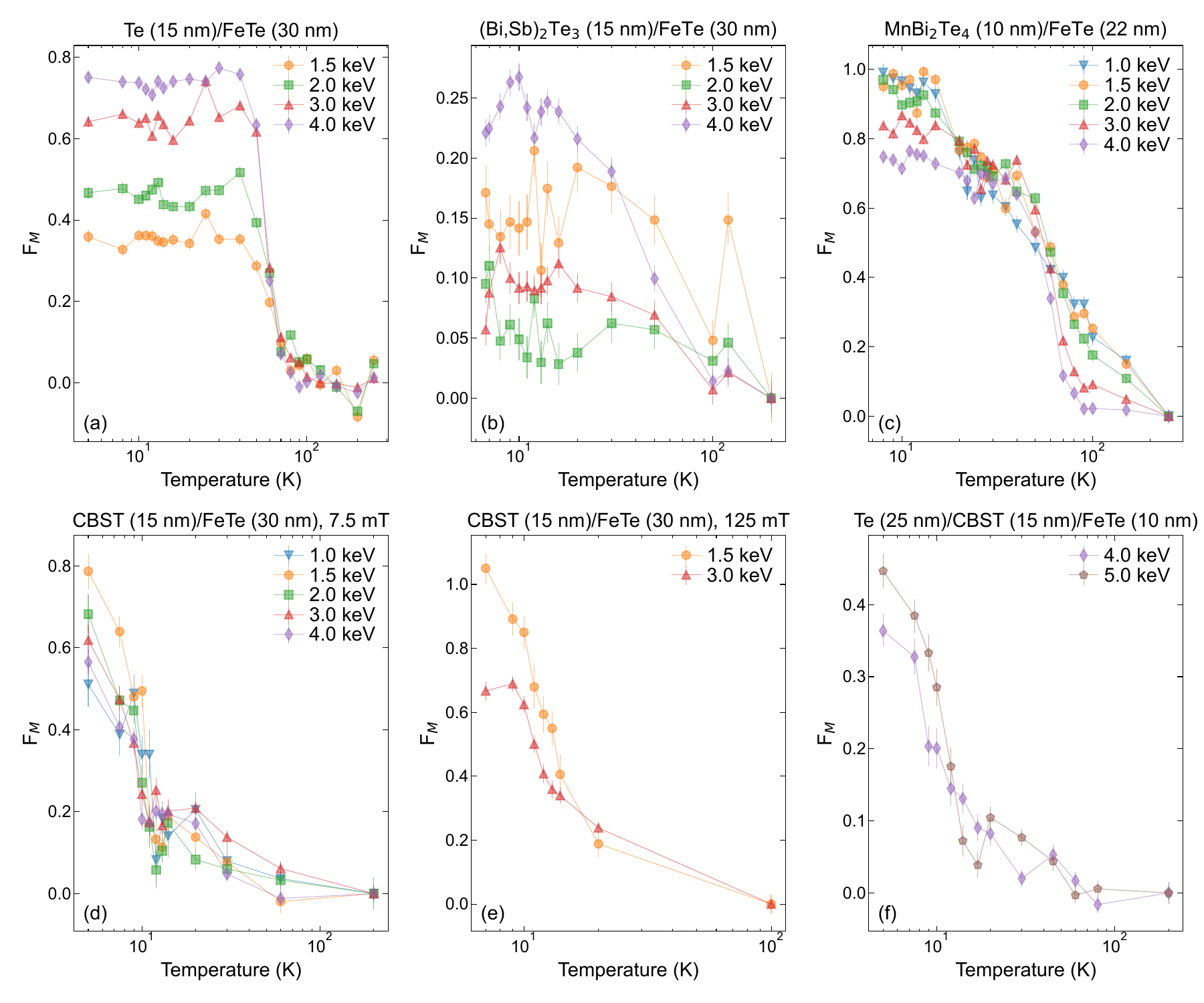}
    \caption{(a) Log-scale temperature- and energy-dependent  $F_M$ values extracted from LE-$\mu$SR measurements for Te (15 nm)/FeTe (30 nm) at 7.5 mT, (b) BST (15 nm)/FeTe (30 nm) at 7.5 mT, (c) MBT (10 nm)/FeTe at 7.5 mT, (d) CBST (15 nm)/FeTe (30 nm) at 7.5 mT, (e) CBST (15 nm)/FeTe (30 nm) at 125 mT, and (f) Te (25 nm)/CBST (15 nm)/FeTe (10 nm) at 7.5 mT. Error bars represent $\pm$1 standard deviation.
    }
    \label{fig:MuonFMFullLog}
\end{figure*}

\begin{figure*}
    \centering
    \includegraphics[width=1.0\textwidth]{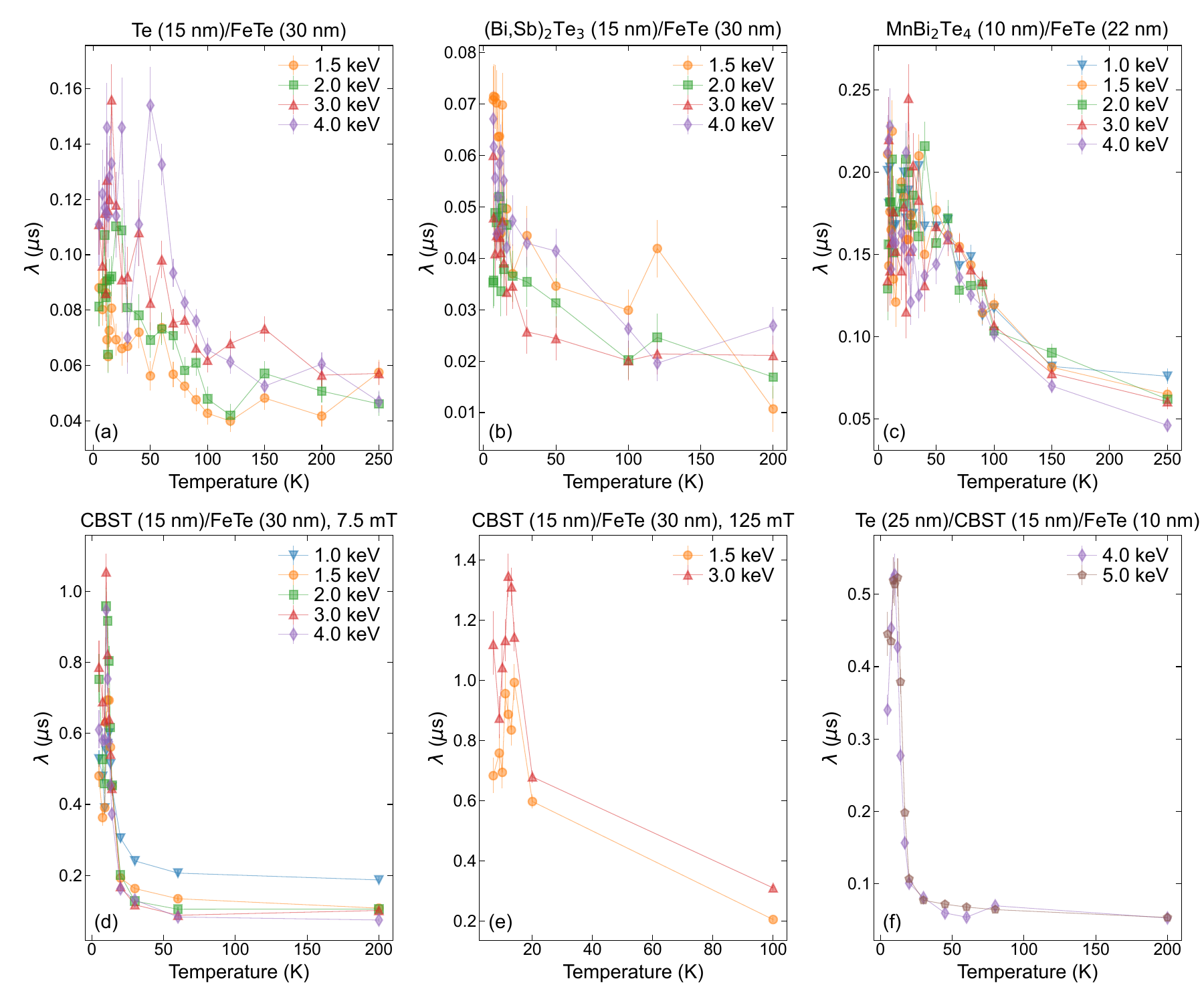}
    \caption{(a) Temperature- and energy-dependent $\lambda$ values extracted from LE-$\mu$SR measurements for Te (15 nm)/FeTe (30 nm) at 7.5 mT, (b) BST (15 nm)/FeTe (30 nm) at 7.5 mT, (c) MBT (10 nm)/FeTe at 7.5 mT, (d) CBST (15 nm)/FeTe (30 nm) at 7.5 mT, (e) CBST (15 nm)/FeTe (30 nm) at 125 mT, and (f) Te (25 nm)/CBST (15 nm)/FeTe (10 nm) at 7.5 mT. Error bars represent $\pm$1 standard deviation.
    }
    \label{fig:MuonLambdaFull}
\end{figure*}

\begin{figure*}
    \centering
    \includegraphics[width=1.0\textwidth]{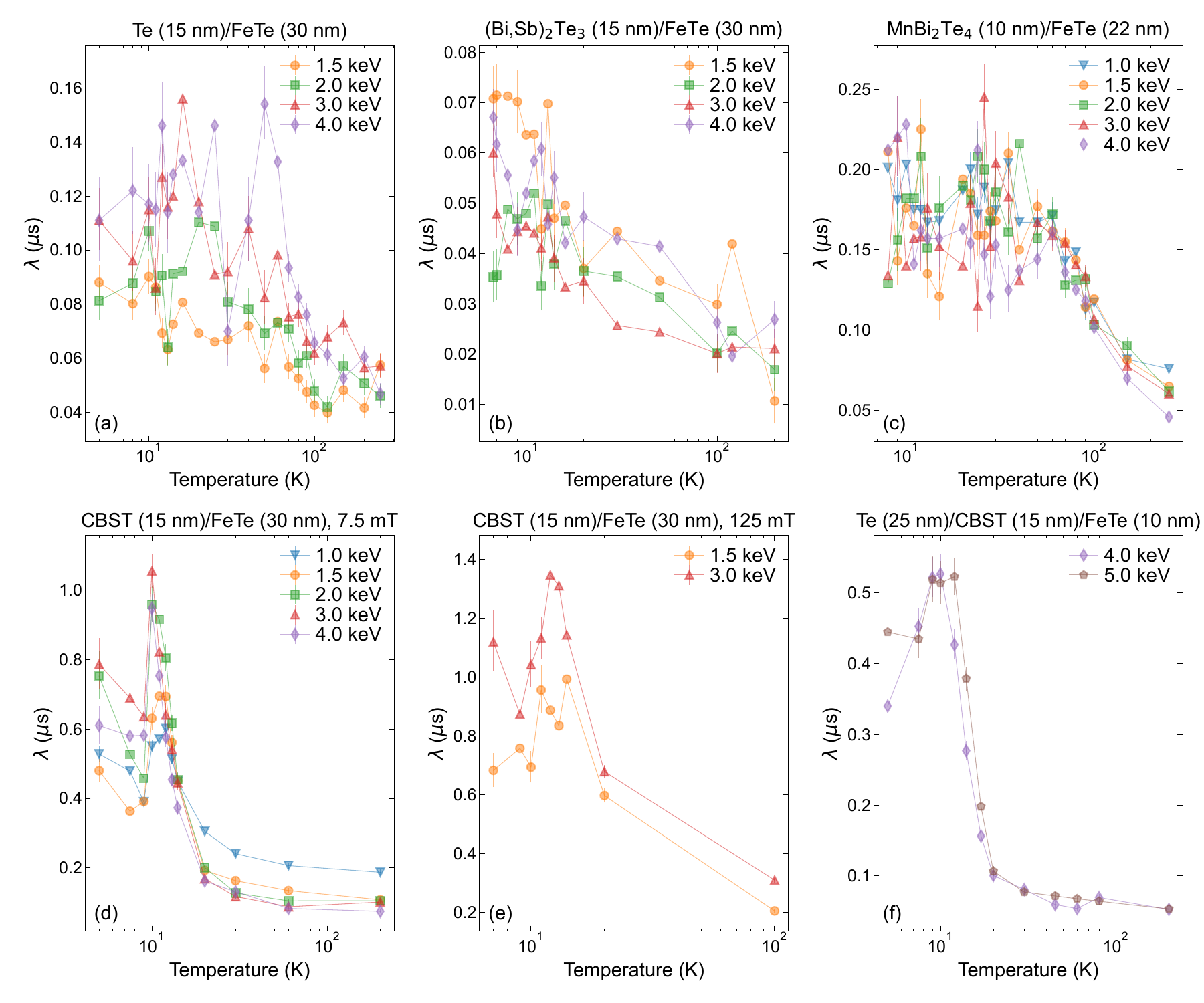}
    \caption{(a) Log-scale temperature- and energy-dependent $\lambda$ values extracted from LE-$\mu$SR measurements for Te (15 nm)/FeTe (30 nm) at 7.5 mT, (b) BST (15 nm)/FeTe (30 nm) at 7.5 mT, (c) MBT (10 nm)/FeTe at 7.5 mT, (d) CBST (15 nm)/FeTe (30 nm) at 7.5 mT, (e) CBST (15 nm)/FeTe (30 nm) at 125 mT, and (f) Te (25 nm)/CBST (15 nm)/FeTe (10 nm) at 7.5 mT. Error bars represent $\pm$1 standard deviation.
    }
    \label{fig:MuonLambdaFullLog}
\end{figure*}

\begin{figure*}
    \centering
    \includegraphics[width=1.0\textwidth]{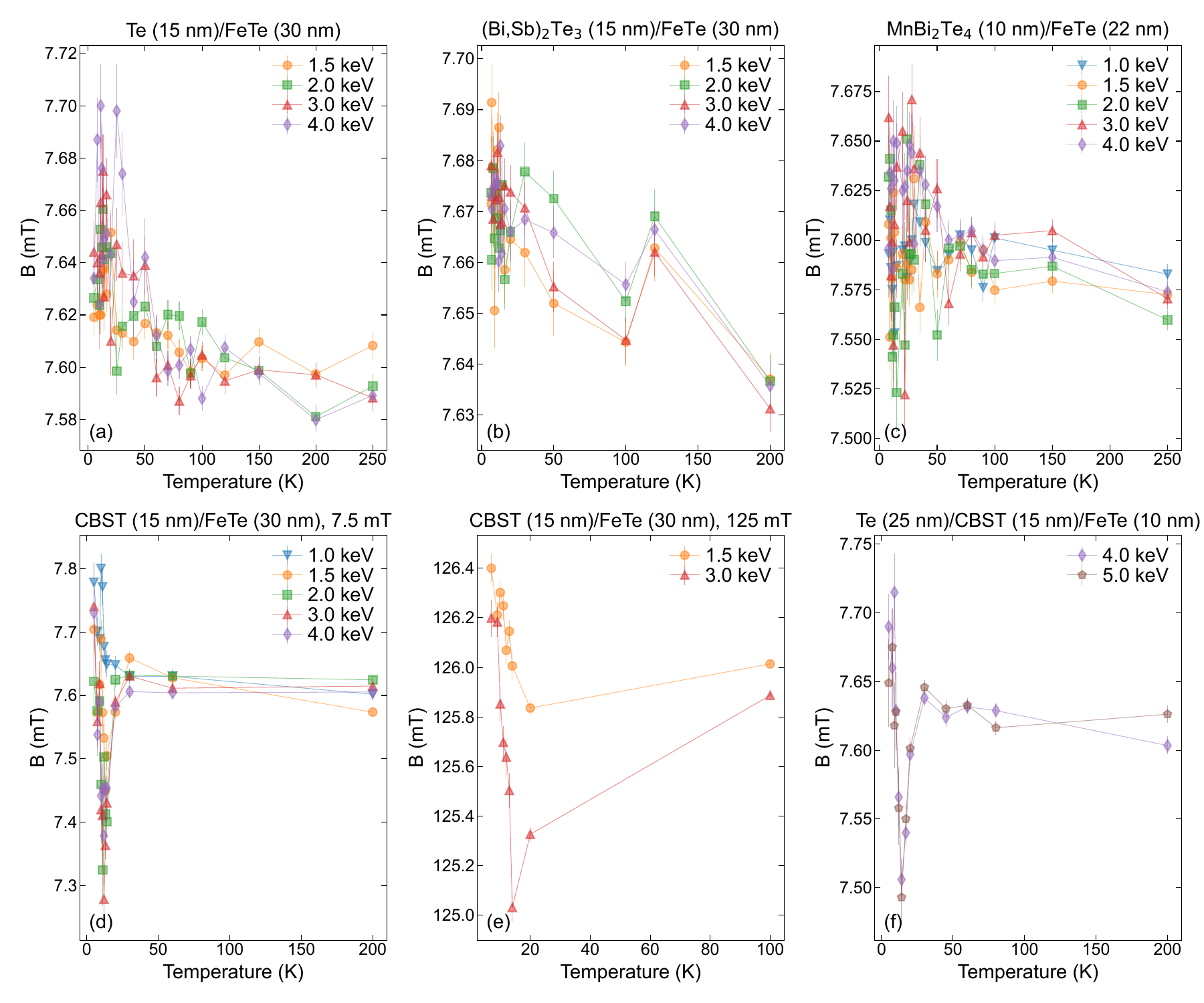}
    \caption{(a) Temperature- and energy-dependent magnetic field (B) values extracted from LE-$\mu$SR measurements for Te (15 nm)/FeTe (30 nm) at 7.5 mT, (b) BST (15 nm)/FeTe (30 nm) at 7.5 mT, (c) MBT (10 nm)/FeTe at 7.5 mT, (d) CBST (15 nm)/FeTe (30 nm) at 7.5 mT, (e) CBST (15 nm)/FeTe (30 nm) at 125 mT, and (f) Te (25 nm)/CBST (15 nm)/FeTe (10 nm) at 7.5 mT. Error bars represent $\pm$1 standard deviation.
    }
    \label{fig:MuonBFull}
\end{figure*}

\begin{figure*}
    \centering
    \includegraphics[width=1.0\textwidth]{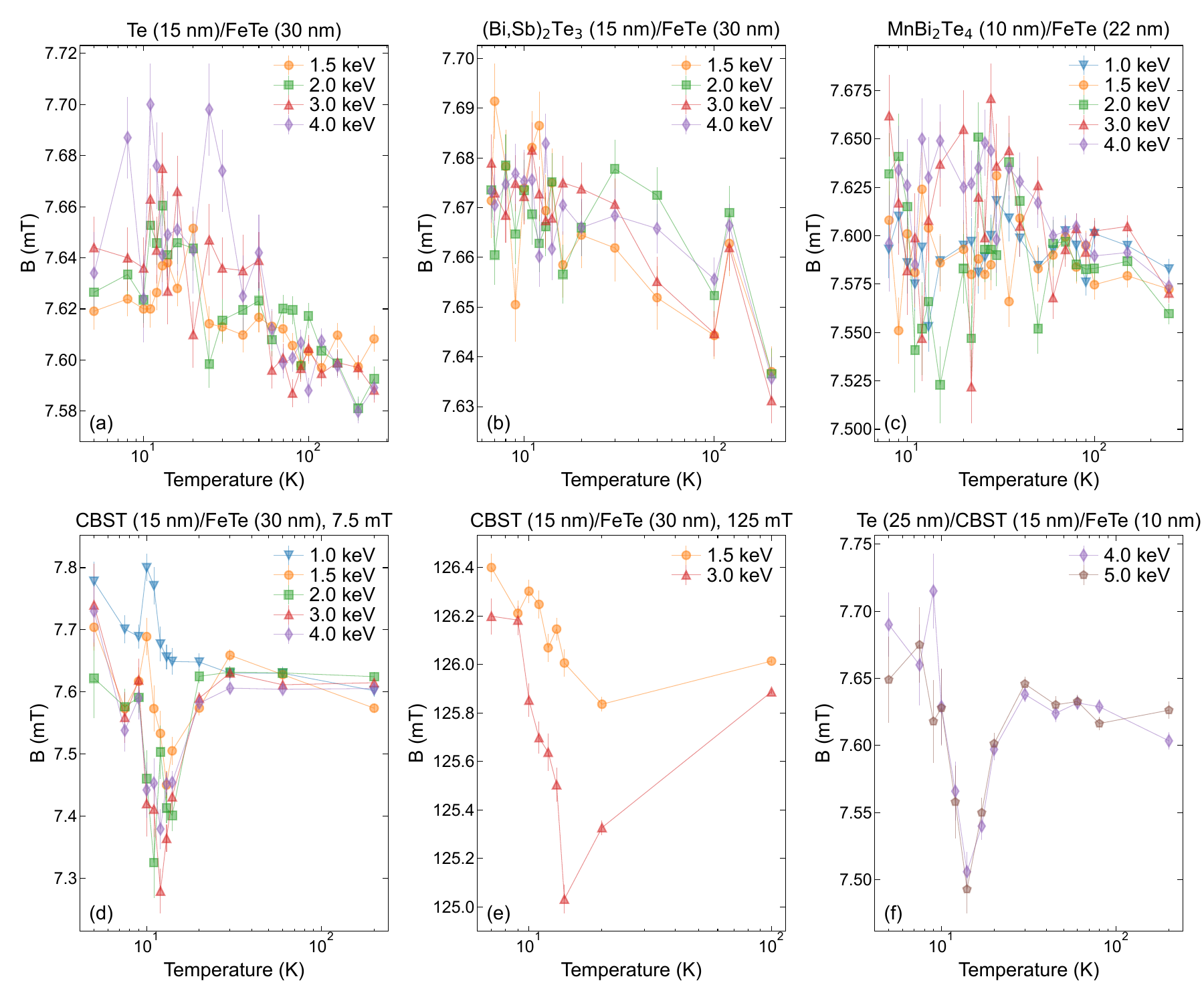}
    \caption{(a) Log-scale temperature- and energy-dependent magnetic field (B) values extracted from LE-$\mu$SR measurements for Te (15 nm)/FeTe (30 nm) at 7.5 mT, (b) BST (15 nm)/FeTe (30 nm) at 7.5 mT, (c) MBT (10 nm)/FeTe at 7.5 mT, (d) CBST (15 nm)/FeTe (30 nm) at 7.5 mT, (e) CBST (15 nm)/FeTe (30 nm) at 125 mT, and (f) Te (25 nm)/CBST (15 nm)/FeTe (10 nm) at 7.5 mT. Error bars represent $\pm$1 standard deviation.
    }
    \label{fig:MuonBFullLog}
\end{figure*}

\begin{figure*}
    \centering
    \includegraphics[width=1.0\textwidth]{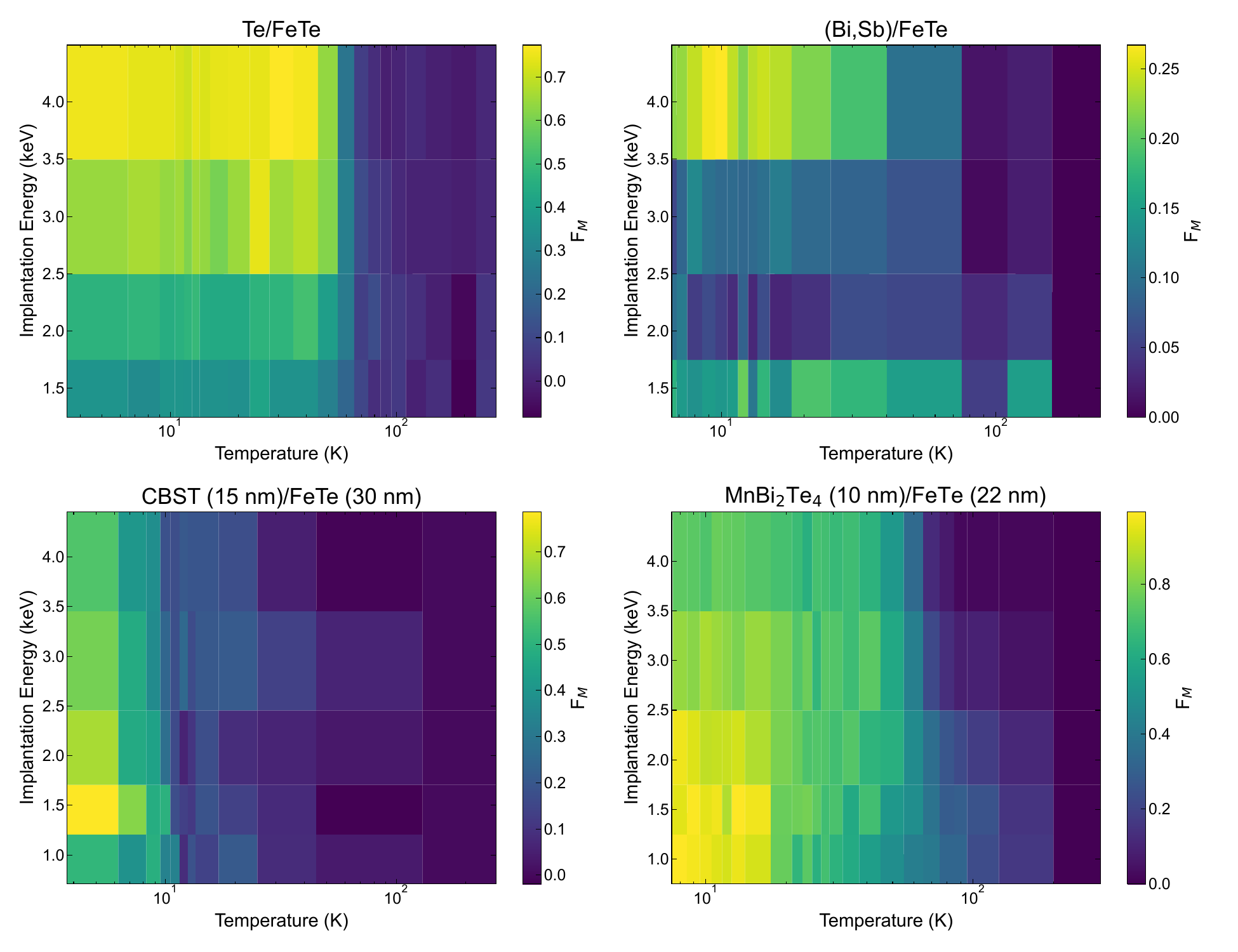}
    \caption{Temperature- and energy-dependent  $F_M$ values extracted from LE-$\mu$SR measurements for Te (15 nm)/FeTe (30 nm), (b) BST (15 nm)/FeTe (30 nm), (c) MBT (10 nm)/FeTe, and (d) CBST (15 nm)/FeTe (30 nm) bilayers. All measurements shown were formed in an applied external magnetic field of approximately 7.5 mT.
    }
    \label{fig:MuonFM_ColorPlot}
\end{figure*}

\subsection{\texorpdfstring{Derivative of $F_M$}{Derivative of Magnetic Volume Fraction}}

\begin{figure*}
    \centering
    \includegraphics[width=1.0\textwidth]{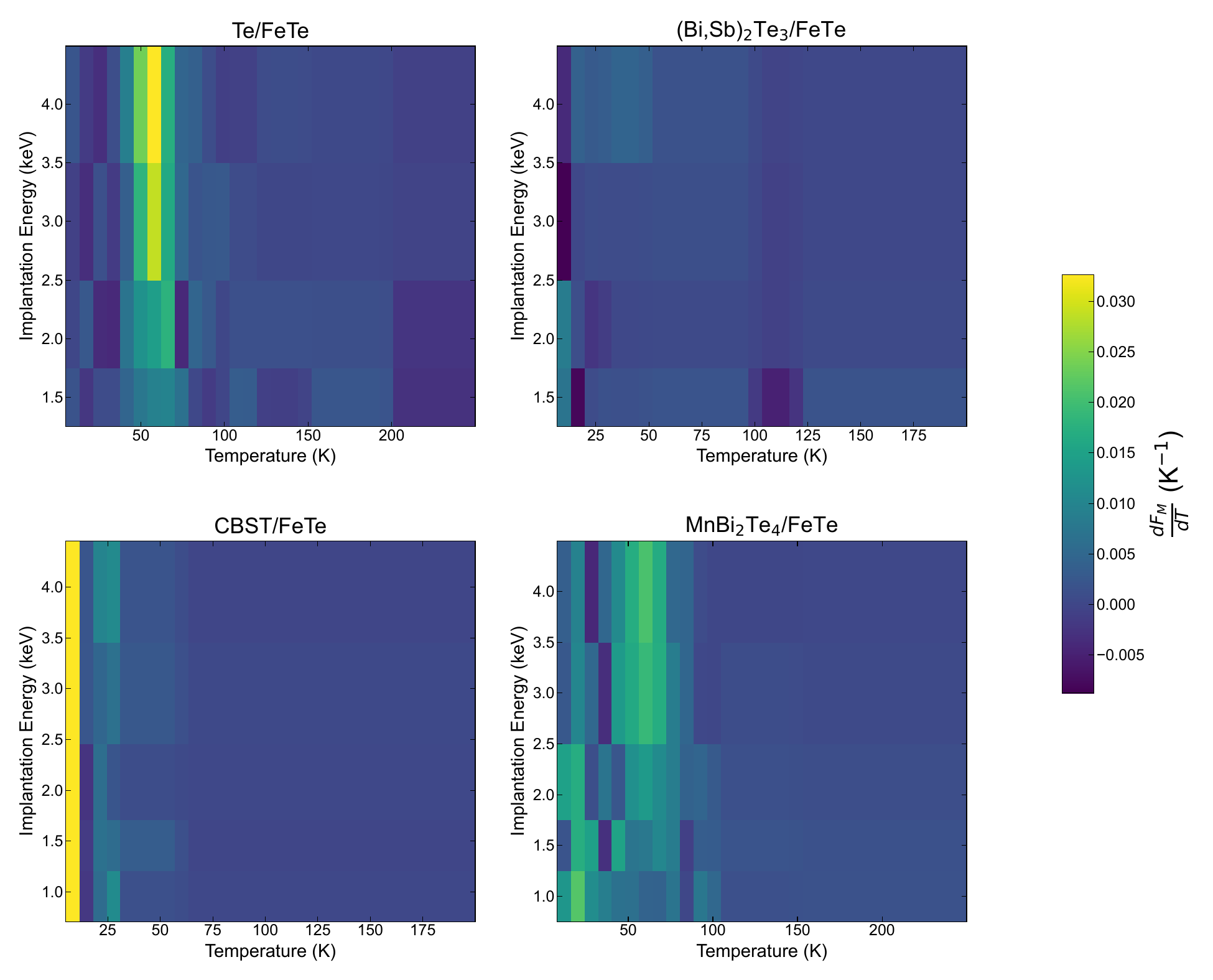}
    \caption{Temperature- and energy-dependent  $F_M$ values extracted from LE-$\mu$SR measurements for Te (15 nm)/FeTe (30 nm), BST (15 nm)/FeTe (30 nm), MBT (10 nm)/FeTe, and CBST (15 nm)/FeTe (30 nm) bilayers. All measurements shown were formed in an applied external magnetic field of approximately 7.5 mT. Data were interpolated onto a 1 K grid, the derivative was calculated, and then data were averaged into bins.
    }
    \label{fig:FMDerivative}
\end{figure*}

To aid in the visualization of magnetic phase transitions, Figure \ref{fig:FMDerivative} shows the derivative of $F_M$ for the Te (15 nm)/FeTe (30 nm), BST (15 nm)/FeTe (30 nm), CBST (15 nm)/FeTe (30 nm), and MBT (10 nm)/FeTe (22 nm) samples. To create this visualization, the data were interpolated onto a regular grid with approximately 1 K temperature spacing, a derivative was taken, and then the derivative was averaged into the displayed temperature bins. A single transition is visible in the Te/FeTe sample, appearing as a single bright feature which is strongest at higher implantation energies. The BST is largely devoid of strong features. The CBST/FeTe data is similar to the BST/FeTe except for an extremely large peak at in the lowest temperature bin. Lastly, the MBT sample is quite complex. The FeTe antiferromagnetic transition appears strongly at higher energies, as in the Te/FeTe sample, but is slightly weaker and broader. On the other hand, the lowest implantation energies show the gradual increase in  $F_M$ at higher temperature, followed by the brighter feature at approximately 20 K - 25 K which represents that antiferromagnetic ordering of MBT.

\subsection{Alternative Fit to PNR Data}

As with many scattering techniques, neutron reflectometry modeling does not provide a unique solution to the real-space magnetic and compositional layer structure, and we must take care when constructing a model to encode everything we know about the sample, including the thickness and composition of each layer deposited during sample creation and the outputs of other characterization techniques that provide constraints on the total magnetization.  Even after taking into account all our existing knowledge of the sample, there are frequently a few possible models that can fit the data, and we examine such alternatives.  For example, a model which allowed a net magnetization in the low-density FeTe transitional growth region yielded an improved fit (lower chi-squared measure) but with a negative magnetization in this layer at 1~T, which violates reasonable physical constraints (we expect the magnetization to be mostly aligned with a field that large, not anti-aligned, based on previous observations of the magnetic susceptibility of FeTe.) It is worth noting that even  in this non-physical model, the CBST and bulk FeTe layers remained positively magnetized, agreeing with the results of the ``best-fit’’ model we have presented above (which does not violate any physical constraints.)

\begin{figure*}
    \centering
    \includegraphics[width=1.0\textwidth]{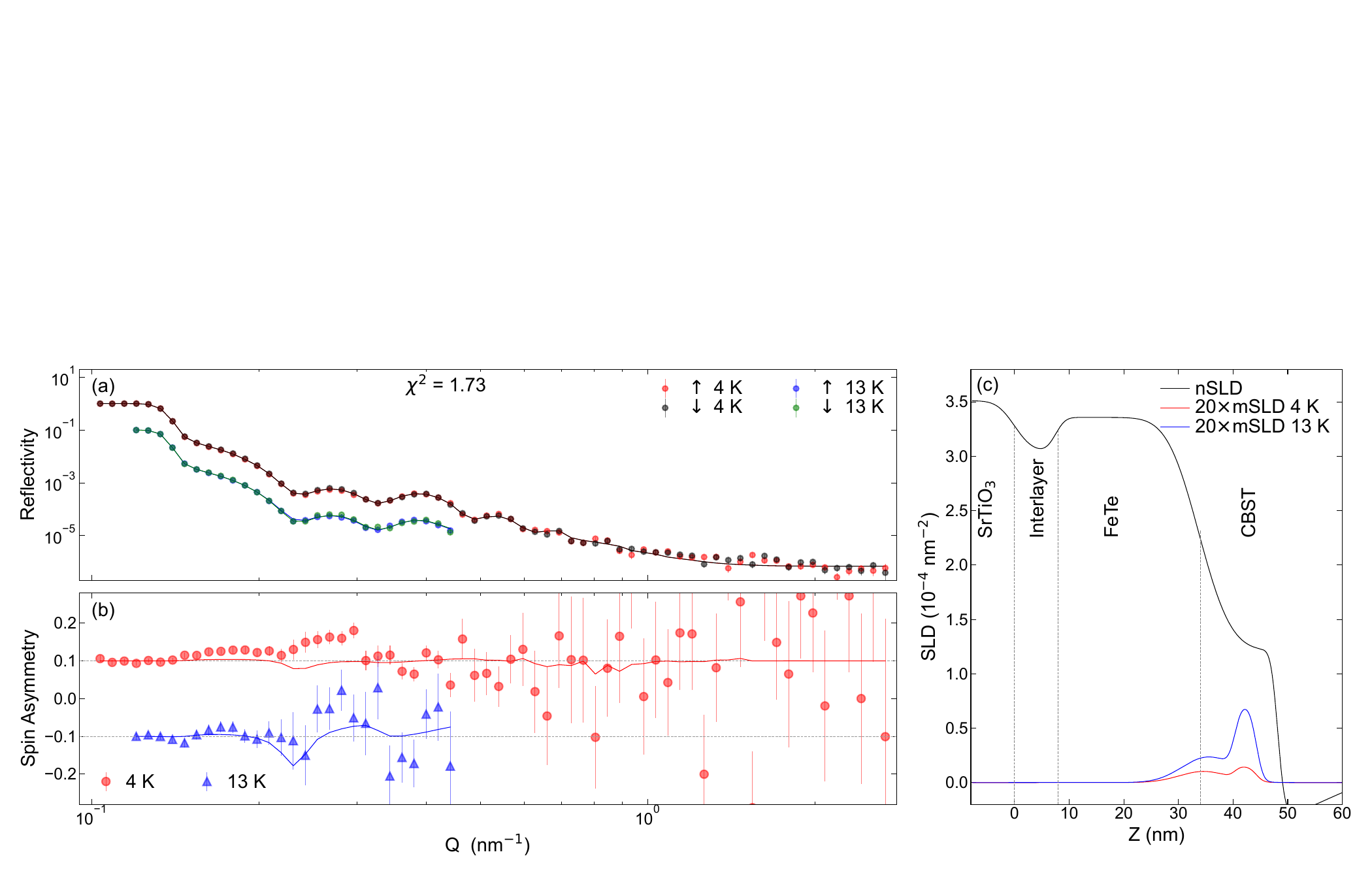}
    \caption{(a) Spin-dependent neutron reflectivities as a function of Q, alongside theoretical fits. Curves offset for visual clarity. (b) Spin-asymmetry and theoretical fits derived from the reflectivity curves shown in \textit{a}. (c) The optimized model used to fit the data, with all nonzero magnetic SLD confined either within the CBST layer or the CBST/FeTe interface.
    }
    \label{fig:BadPNR}
\end{figure*}

For comparison, we present below the best physically reasonable alternative model we have identified, in which all of the magnetization is confined to the CBST layer and the CBST/FeTe interface. This model is shown in Figure \ref{fig:BadPNR}, where it can be seen that the low-Q spin splitting is very poorly described, especially in the superconducting state.

In fitting the data, we used both simultaneous coupled models fitting both the 13 K and 4 K data simultaneously, as well as independent fits of each temperature. The results were effectively identical, although we reproduce the best-fit coupled model for completeness below in Figure \ref{fig:CoupledFit}

\begin{figure*}
    \centering
    \includegraphics[width=1.0\textwidth]{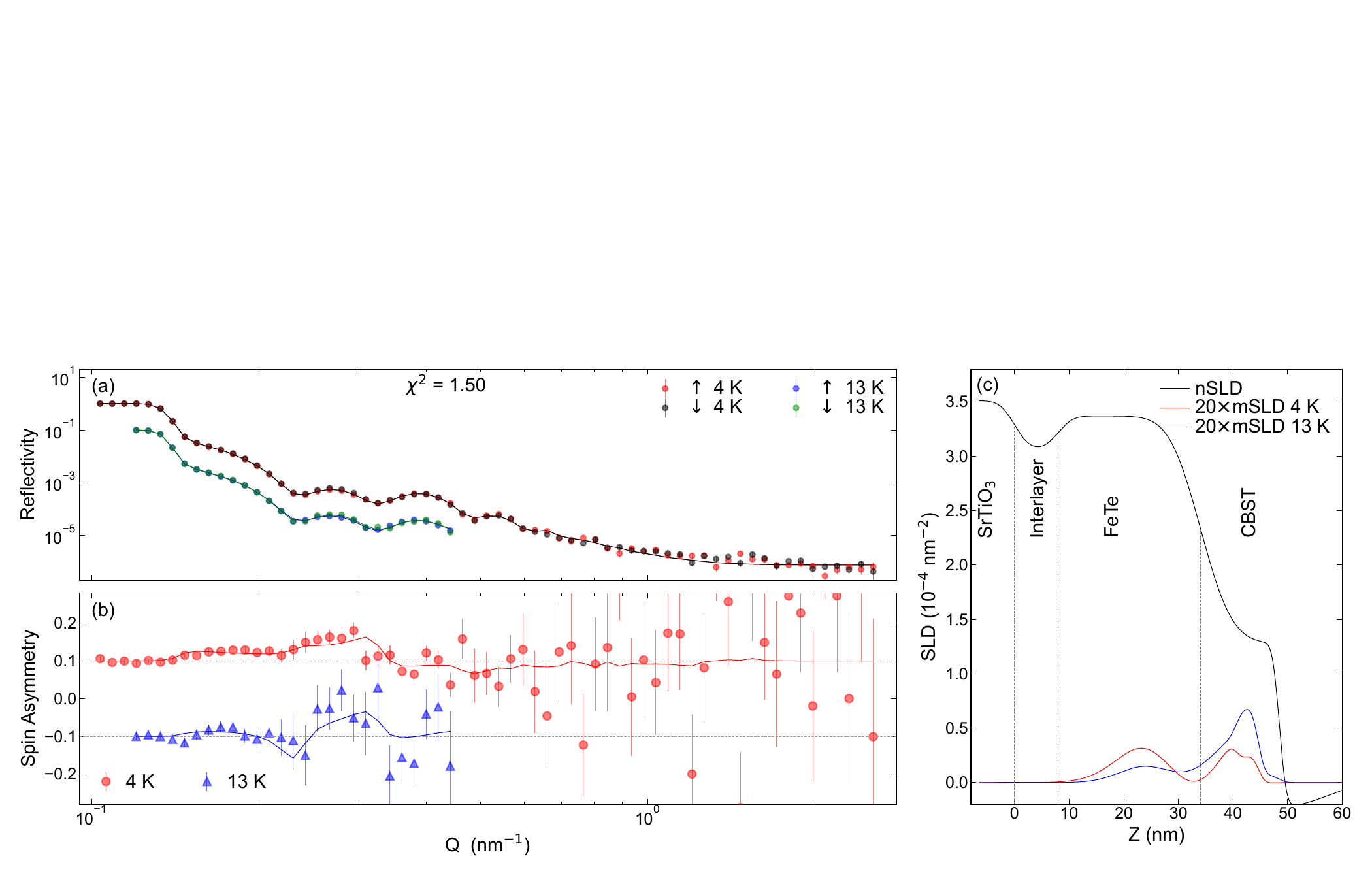}
    \caption{(a) Spin-dependent neutron reflectivities as a function of Q, alongside theoretical fits. Curves offset for visual clarity. (b) Spin-asymmetry and theoretical fits derived from the reflectivity curves shown in \textit{a}. (c) The optimized model used to fit the data, with all nonzero magnetic SLD confined either within the CBST layer or the CBST/FeTe interface.
    }
    \label{fig:CoupledFit}
\end{figure*}

\subsection{Representative Transport Measurements}

\begin{figure*}
    \centering
    \includegraphics[width=1.0\textwidth]{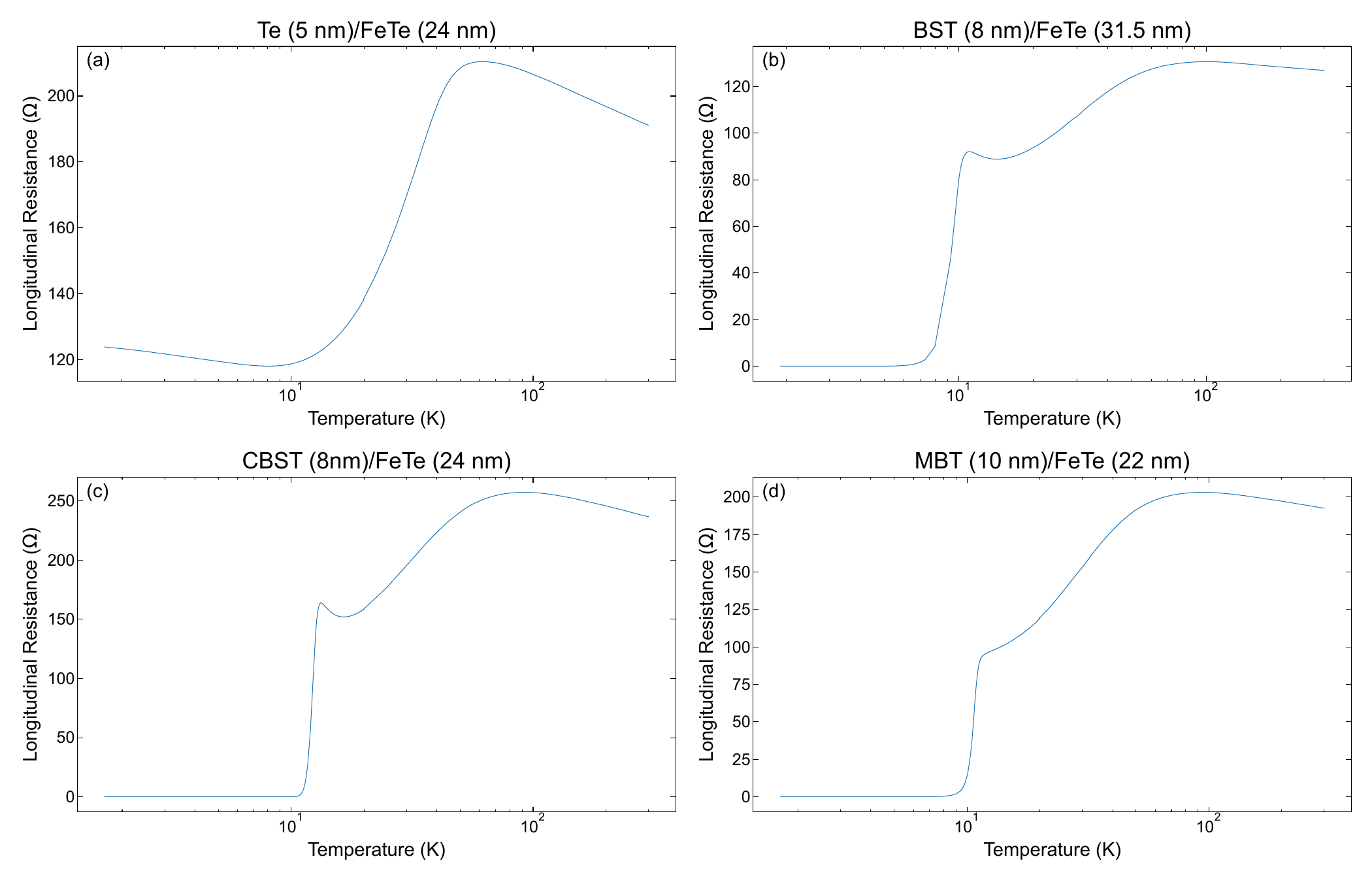}
    \caption{Longitudinal resistance vs. temperature for (a) Te (5 nm)/FeTe (24 nm), (b) BST (8 nm)/FeTe (31.5 nm), (c) CBST (8 nm)/FeTe (24 nm), and (d) MBT (10 nm)/FeTe (22 nm).
    }
    \label{fig:transport}
\end{figure*}

In Figure \ref{fig:transport}, we show representative longitudinal resistance (frequently referred to as R$_{xx}$) curves as a function of temperature for every type of bilayer used in this study. It is well known that the interfacial superconductivity in TI/FeTe bilayers is extraordinarily robust, spanning an extremely wide range of thicknesses for both TI and FeTe layers, as well as surviving extended film aging. As expected, Te-capped FeTe shows no superconducting transition.

\FloatBarrier
%\section*{References}
\twocolumngrid

\bibliography{references}

\end{document}